\documentclass[amsmath,amssymb,prb,twocolumn,superscriptaddress,longbibliography,hypertext]{revtex4-2}

\usepackage{float}
\usepackage{siunitx}
\usepackage{booktabs}
\usepackage{graphicx}
\usepackage{array}
\usepackage{amsthm}
\usepackage{enumerate}
\usepackage[table]{xcolor}
\usepackage{bbm}
\usepackage{color}
\usepackage[normalem]{ulem}
\usepackage{multirow}
\RequirePackage[colorlinks=true,linkcolor=blue]{hyperref}
\usepackage{braket}

\newcommand{\Appref}[1]{App.~\ref{#1}}
\newcommand{\Secref}[1]{Sec.~\ref{#1}}
\newcommand{\Figref}[1]{Fig.~\ref{#1}}
\newcommand{\Tabref}[1]{Tab.~\ref{#1}}
\newcommand{\Eqref}[1]{Eq.~\eqref{#1}}

\newcommand{\dipc}{Donostia International Physics Center (DIPC), E-20018, Donostia-San Sebasti\'an, Spain}
\newcommand{\ikerbasque}{IKERBASQUE, Basque Foundation for Science, E-48013, Bilbao, Spain}

\newcommand{\ketbra}[2]{\ket{#1} \bra{#2}}
\newcommand{\proj}[1]{\ket{#1}\bra{#1}}
\newcommand{\up}{\hspace{-0.2 pt}\uparrow}
\newcommand{\dn}{\hspace{-0.2 pt}\downarrow}

\newcommand{\br}{\mathbf{r}}
\newcommand{\bI}{\mathbf{I}}
\newcommand{\bK}{\mathbf{K}}
\newcommand{\bR}{\mathbf{R}}
\newcommand{\bS}{\mathbf{S}}
\newcommand{\id}{\mathbbm{1}}

\newcommand{\carbon}{$^{13}$C}
\newcommand{\hydrogen}{$^{1}$H}
\newcommand{\orca}{\textsc{Orca}}
\newcommand{\siesta}{\textsc{Siesta}}
\newcommand{\hubbard}{\textsc{Hubbard}}
\newcommand{\pol}{$\pi$-spin polarization}
\newcommand{\orcaxc}{\orca\ using B3LYP \cite{dataset}}

\newcommand{\hfispinpolplot}{Red and blue blobs depict negative and positive values, respectively. The area of the circle is proportional to the magnitude of the respective quantity at that site. As a reference scale, the gray blobs represent 100 MHz (left) and one electron spin-1/2 (right), respectively.}

\newcommand{\ie}{\emph{i.e.}}
\newcommand{\eg}{\emph{e.g.}}

\begin{document}

\title{Hyperfine interactions in open-shell planar $sp^2$-carbon nanostructures}

\author{Sanghita Sengupta}
\affiliation{\dipc}
\author{Thomas Frederiksen}
\affiliation{\dipc}
\affiliation{\ikerbasque}
\author{Geza Giedke}
\affiliation{\dipc}
\affiliation{\ikerbasque}

\date{\today}

\begin{abstract}
    We investigate hyperfine interaction (HFI) using density-functional theory for several open-shell planar $sp^2$-carbon nanostructures displaying $\pi$ magnetism. Our prototype structures include both benzenoid ([$n$]triangulenes and a graphene nanoribbon) as well as non-benzenoid (indene, fluorene, and indene[2,1-b]fluorene) molecules. Our results obtained with \orca\ indicate that isotropic Fermi contact and anisotropic dipolar terms contribute in comparable strength, rendering the HFI markedly anisotropic. We find that the magnitude of HFI in these molecules can reach more than 100 MHz, thereby opening up the possibility of experimental detection via methods such as electron spin resonance-scanning tunneling microscopy (ESR-STM). Using these results, we obtain empirical models based on $\pi$-spin polarizations at carbon sites. These are defined by generic $sp^{2}$ HFI fit parameters which are derived by matching the computed HFI couplings to $\pi$-spin polarizations computed with methods such as \textsc{Orca}, \textsc{Siesta}, or mean-field Hubbard (MFH) models. This approach successfully describes the Fermi contact and dipolar contributions for $^{13}$C and $^{1}$H nuclei. These fit parameters allow to obtain hyperfine tensors for large systems where existing methodology is not suitable or computationally too expensive. As an example, we show how HFI scales with system size in [$n$]triangulenes for large $n$ using MFH. We also discuss some implications of HFI for electron-spin decoherence and for coherent nuclear dynamics.
\end{abstract}

\maketitle

\section{Introduction}
\label{sec:Intro}
Magnetism in graphene-related structures is of immense interest because of promising technological applications such as spintronics, where the electronic spin degree of freedom is manipulated instead of the charge \cite{dassharma, Han2014}. Electronic spin not only serves as a crucial ingredient in spin-based devices but is of paramount importance in the field of quantum computation \cite{loss}. The building blocks for these devices require an understanding of the transport mechanisms related to non-equilibrium spin dynamics as well as spin relaxation. In solids, the presence of spin-orbit coupling (SOC) and hyperfine interactions (HFI) renders the spin a non-conserved quantity, thereby making it necessary to understand how these interactions affect the electronic spins \cite{elliot, Yafet1963, dyakonov1972spin, dassharma, WU201061, hanson}.

The intrinsic SOC in graphene-based materials is well-known to generate a gap of the order of $\sim$ 100-200 MHz \cite{yao, min} and, for localized and energetically isolated electronic states, it is expected to contribute only weakly to spin decoherence, as was studied in detail for quantum dots \cite {golovach04}. HFI, on the other hand, has been supposed to be similarly small in graphene due to the low natural abundance (1\%) of spinful \carbon\ nuclei and the $\pi$-electron character of the low-energy electronic states \cite{TrBuLo.07.Spinqubitsgraphene,Fischer09}.
For the hydrocarbon nanostructures of interest here, a full quantitative understanding of HFI is still being sought, in particular due to the presence of (always spinful) \hydrogen\ nuclei at the edges. A quantitative knowledge of HFIs will be beneficial in understanding the main mechanisms for spin relaxation and spin decoherence, similar to that of localized electrons in quantum dots \cite{hanson, glazman, khaetskii, merkulov, coish}. HFIs also aid in our comprehension of the electronic structure of the materials.
Well-established resonance techniques such as electron spin resonance (ESR), electron paramagnetic resonance (EPR), and nuclear magnetic resonance (NMR) \cite{abragam2012electron, nla.cat-vn1417077} have long been used to study spin-related properties of materials and have recently been implemented with real-space atomic resolution in scanning tunneling microscopy (ESR-STM) \cite{BaPaCh.15.Electronparamagneticresonance, wilke} and in atomic force microscopy (ESR-AFM) \cite{repp22}.

The main focus of this work is the computation of hyperfine tensors for a series of magnetic molecules belonging to the set of planar $sp^2$-carbon nanostructures. Specifically, we consider molecules that are referred to as $\pi$ radicals as they exhibit unpaired spin densities in the $\pi$ orbitals.
In recent years such nanographenes and related molecules have garnered enormous attention due to their novel physical properties related to magnetism as well as unconventional topological phases of matter \cite{jiang2007first, rossier, Yazyev_2010, Mishra2020, enoki, OtFr.22.Carbonbasednanostructures, Mishra2021, doi:10.1021/jacs.9b09212, Li2019, Li2021, frac_mishra}. Thus these molecules are versatile candidates for potential devices and applications in spintronics \cite{dassharma, Han2014} and quantum computation \cite{Heinrich2021}.

Our theoretical work is built on first-principles calculations based on density-functional theory (DFT) as implemented in the all-electron code \orca\ \cite{orca}.
This general-purpose quantum chemistry code, based on Gaussian basis functions, can provide accurate spectroscopic properties of large open-shell molecules. In fact, obtaining ESR and NMR spectra from DFT linear-response methods has been central from the beginning of its code development \cite{orca}. 
While several alternative DFT-based approaches have been applied in related contexts for solids \cite{GaFyKa.08.Abinitiosupercell,wilke,Philippopoulos2020, ShDiGu.21.Trendshyperfineinteractions} and isolated molecules \cite{Ya.08.HyperfineInteractionsGraphene}, we consider \orca\ a state-of-the-art approach to the problem at hand.

In addition to reporting the DFT-calculated HFIs for planar $sp^2$-carbon  nanostructures, we also lay out a parametrization procedure which relates the HFIs to the atom-resolved \pol\ at the carbon sites. Similar parametrization models have been routinely used in $\pi$ radicals \cite{McCh.58.TheoryIsotropicHyperfine, KaFr.61.TheoreticalInterpretationCarbon13, weil_bolton, Ya.08.HyperfineInteractionsGraphene} and our work extends it to nanographenes not only for the \carbon\ but also \hydrogen\  nuclei, which are important since they typically represent the largest number of nuclear spins in hydrocarbon nanostructures with natural abundance of \carbon. We thus obtain generic hydrocarbon $sp^2$ fit parameters which are sufficient to describe the magnitudes of hyperfine couplings in all $sp^2$-carbon structures once the \pol s at the carbon sites are known. 
We show that this method works well for a range of planar structures in which the magnetic properties arise primarily from the $\pi$ electrons.

The benefit of this fitting procedure is the possibility -- as also investigated in this work -- to then use other tools, such as \siesta\ DFT \cite{soler} and mean-field Hubbard (MFH) models \cite{rossier, Yazyev_2010, SaPaGi.22.SpinPolarizingElectron, dipc_hubbard}, which are known to give satisfactory results for \pol\ at the carbon sites \cite{OtFr.22.Carbonbasednanostructures}. The unique feature of this parametrization procedure lies in its generality.
Any electronic-structure method that can reliably compute \pol, can then be used to efficiently predict hyperfine couplings, given appropriate fit parameters. This is opposed to a direct evaluation of hyperfine couplings in the real-space integral formalism which relies on an accurate description of the spin density near the nuclei,
and thus requires special attention in electronic structure calculations for the valence electrons only: (i) a correct description of the wave functions near the nuclei
as developed within the projector augmented wave (PAW) formalism \cite{blochlhfi}, and to (ii) corrections from polarization of core-electrons \cite{yazyevvasp}.

Additionally, we point out that the strength of the parametrization technique is aptly realized when combined with the MFH approach. The simplistic approach with a single $\pi$ orbital on the carbon atom and only considering onsite Coulomb repulsion $U$, results in computational efficiency, which in turn enables predictions of hyperfine tensors in larger system sizes  where existing methodology is not suitable or computationally too expensive.

The remainder of the article is organized as follows: in \Secref{sec:2.1-HFI}, we introduce the general formulation of hyperfine interaction. We present the computational procedure along with an example of a model molecule (charged anthracenes) in \Secref{sec:2.3-procedure}. The main result of this work, the HFIs of a series of $sp^2$-carbon nanostructures, viz., [$n$]triangulenes, an armchair graphene nanoribbon, and small non-benzenoid molecules, are given in \Secref{sec:HFInanographene}. Next, we describe the parametrization procedure along with empirical models in \Secref{sec:3-fitting-hfi} and lay out the generic $sp^2$ fit parameters. As a check for accuracy we show how these fit parameters perform for known experimental HFIs in case of anthracene molecules. In \Secref{sec:large-molecules-MFH}, we use MFH and apply the fit to study the scaling of HFI in $[n]$triangulenes for large $n$. And in \Secref{sec:application}, we discuss few exemplary applications of hyperfine tensor such as estimating spin-qubit dephasing times or generating entanglement between distant nuclear spins. Finally, we summarize our results in \Secref{sec:discussion} and discuss the possible experimental consequences of measuring these hyperfine couplings.

\section{General formulation of the hyperfine interaction}
\label{sec:2.1-HFI}

We begin this section by providing a general overview of hyperfine interactions and its relation with the electronic spin density. Here, we consider a single multi-electron energy eigenstate of fixed spin and wavefunction which is energetically separated by a gap much larger than the HFI energy of $\sim 100$ MHz from other states. This ensures that no transition involving a change in electronic spin $S$ or the wavefunction need to be considered. The spin density, which typically contains positive and negative parts, is computed for the $S_z=S$ state and assumed to be identical for all spin states up to spin rotations. The formulas given below are specialized to the non-relativistic limit \cite{blugel, pyykko, Philippopoulos2020} for light nuclei such as carbon and hydrogen and to the case of $I=1/2$ nuclear spins for \carbon\ and \hydrogen.

For a nucleus located at $\bR_N$, HFI is usually written in terms of a symmetric $3\times3$ hyperfine tensor $A_N$ such that
\begin{equation}\label{hamhfi}
H_{\mathrm{hfi}}= \bS\cdot A_N \bI_N, 
\end{equation}
where $\bS$ and $\bI_N$ represent the electron and nuclear spin operators, respectively \cite{abragam1961principles, Stoneham2001}. 
Much of this work is focused on computing the hyperfine coupling matrix $A_{N}$ for \carbon\ and \hydrogen\ nuclei. 
The hyperfine tensor $A_{N}$ is the sum of the isotropic (Fermi contact) and anisotropic (dipolar) contributions \cite{abragam1961principles, Stoneham2001, orca, Ya.08.HyperfineInteractionsGraphene}
\begin{equation}\label{Atotal}
A_{N,\mu\nu} = A^\mathrm{iso}_{N} \delta_{\mu\nu} + A^{\mathrm{dip}}_{N,\mu\nu}.
\end{equation}

The Fermi contact term represents an isotropic interaction that arises from finite spin density $\rho(\br) = \rho^{\up}(\br) - \rho^{\dn}(\br)$ at the location $\bR_N$ of the nucleus $N$, such that
\begin{equation}\label{fermi}
A^\mathrm{iso}_{N} = \frac{2\mu_{0}}{3S}\gamma_{e}\gamma_{N}\rho(\bR_{N}),
\end{equation}
where $\mu_{0}$ is the magnetic susceptibility of free space, $\gamma_{e} = g_{e}\mu_{e}$ and $\gamma_{N} = g_{N}\mu_{N}$ with electron $g_{e}$ factor $\sim 2.002$ and nuclear $g_{N}^{\mathrm{C}} = 1.404$ (\carbon), $g_{N}^{\mathrm{H}} = 5.585$ (\hydrogen). $\mu_{e}$ and $\mu_{N}$ are the Bohr and nuclear magnetons, respectively \cite{orca, Ya.08.HyperfineInteractionsGraphene}. 

The dipolar term is an anisotropic contribution that describes the magnetic dipole interaction of the magnetic nucleus with the magnetic moment of the electron, written as
\begin{equation}\label{dipolar}
A^{\mathrm{dip}}_{N,\mu\nu} 
= \frac{\mu_{0}\gamma_{e}\gamma_{N}}{4\pi S}\int\frac{\rho(\br + \bR_{N})}{r^{3}}\frac{3r_\mu r_\nu - \delta_{\mu\nu}r^{2}}{r^{2}}d\br,
\end{equation}
with $\mu,\nu=x,y,z$ \cite{orca, Ya.08.HyperfineInteractionsGraphene}. 

We note that, computationally, the spin density will be expressed in terms of (in general non-orthogonal) basis functions $\phi_l$ as $\rho(\br) = \sum_{kl}\rho_{kl} \phi_{k}^{*}(\br)\phi_{l}(\br)$, where the index labels both the site and the orbital character of the basis function and $\rho_{kl}$ represents the spin density matrix in that basis.
This basis enables to associate a certain \pol\ $\Pi_{N}$ to each atomic site $N$
through the so-called Mulliken \pol, defined as 
\begin{equation}\label{mullikenpi}
\Pi_{N} = \sum_{k\in (N,p_z)}\sum_{l}\rho_{kl}\int \phi_{k}^{*}(\br) \phi_{l}(\br) d\br,
\end{equation}
\ie, we sum the terms $(kl)$ of the spin density matrix with $k=(N,p_z)$ over all $l$ and integrate the spatial part which gives rise to the $kl$-th matrix element of the overlap matrix.

We note here that magnetism in graphene is often related to the $\pi$ electrons \cite{rossier, Yazyev_2010, OtFr.22.Carbonbasednanostructures}. 
However, if the electron spin density were solely formed by $\pi$ electrons, it would imply a vanishing Fermi contact interaction in \Eqref{fermi} due to the nodal structure of the $p$-orbitals at the core. This has sometimes been cited to suggest that HFI is weak in graphene structures \cite{TrBuLo.07.Spinqubitsgraphene, Fischer09}. But as already stated in \cite{Ya.08.HyperfineInteractionsGraphene} and \cite{KaFr.61.TheoreticalInterpretationCarbon13}, the core $1s$ and $2s$ orbitals do play a role via the $\sigma$-spin density.

Let us mention that we have also investigated some other relevant interactions for these molecules but found them to be much weaker than the HFIs we consider. These are (i) \emph{orbital} HFI which we found to be $\sim 10^{-3}$ MHz for both \hydrogen\ and \carbon\ nuclei. (ii) \emph{quadrupolar} interactions are absent since all nuclei considered are spin-1/2. (iii) \emph{nuclear-nuclear} dipolar interactions between \carbon\ and \hydrogen\ were found to be $< 10^{-3}$ MHz. Hence we neglect all of these in the following sections.

In the next section, we introduce the computational procedure that we use to calculate the HFIs and in order to check the validity of the method, we will apply the procedure on charged anthracene molecules. These benzenoid hydrocarbon systems serve as prototype molecules for which the \carbon\ and \hydrogen\ isotropic hyperfine couplings are well-measured.

\section{Computational Procedure \& Experimentally studied model molecule}
\label{sec:2.3-procedure}

\begin{table}
\hspace{-1cm}
\begin{minipage}{.2\textwidth}
	\includegraphics[width=4.5cm]{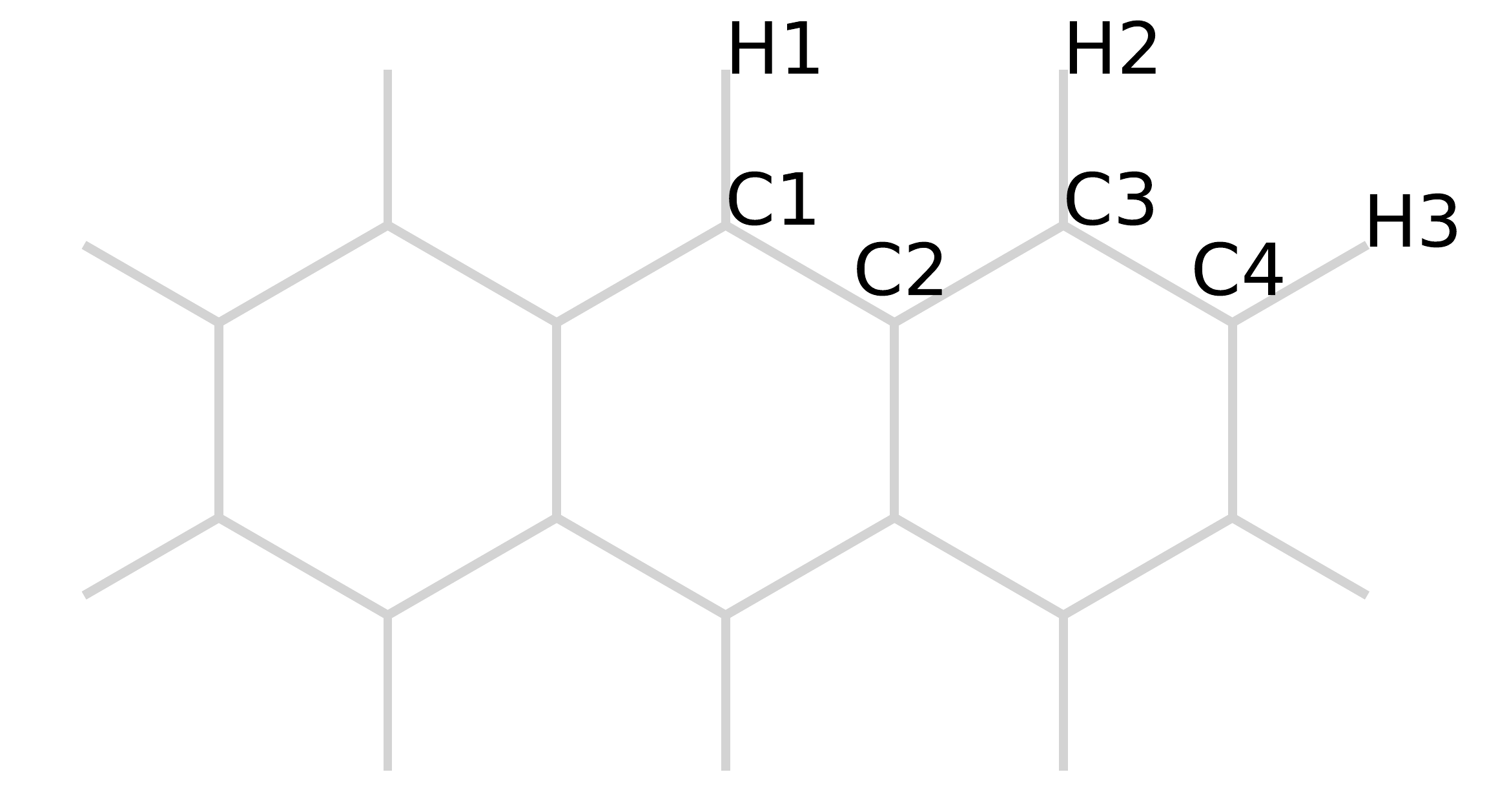}
\end{minipage}
\begin{tabular}{lrrrrrrr}
\toprule
		& \multicolumn{2}{c}{Experiment}& \multicolumn{2}{c}{\orca}	& \multicolumn{3}{c}{Empirical model}\\
$N$		& {+}		& {$-$}				& {+} 		& {$-$}			& {+} 		& {$-$} \\[0.5ex] 
\midrule
C1 		& 23.8		& 24.5				& 21.4		& 20.9			& 22.8		& 23.3 \\
C2		& $-12.6$ 	& $-12.9$			& $-12.8$ 	& $-13.5$		& $-13.4$	& $-13.4$ \\
C3		& {---} 	& 10.0 				& 9.1		& 10.3			& 9.9		& 9.5 \\
C4 		& {$\pm$}1.0 & {$-0.7$}			& $-2.0$	& $-1.8$		& $-1.9$	& $-1.7$ \\
H1 		& $-18.3$	& $-15.0$			& $-18.3$	& $-15.6$		& $-19.1$	& $-19.6$ \\
H2 		& $-8.6$	& $-7.7$			& $-9.4$	& $-9.4$		& $-9.3$	& $-8.9$ \\
H3 		& $-3.9$	& $-4.2$			& $-3.5$ 	& $-4.1$		& $-2.9$	& $-3.0$ \\
\bottomrule
\end{tabular}
\caption{\label{tab:anthracene} \carbon\ and \hydrogen\ isotropic Fermi contact HFIs in charged spin doublet anthracene molecules (in units of MHz). $+$ and $-$ refer to the positive and negative ion anthracenes, respectively.
	The experimental data originate from Refs.~\onlinecite{bolton, weil_bolton}.
	The \orca\ results were computed using EPR-II/B3LYP \cite{dataset}.
	Details for the \emph{empirical model} are given in \Secref{sec:3-fitting-hfi}.}
\end{table}

The first-principles calculations were performed within the \orca\ package which is an all-electron DFT code \cite{orca, dataset}. Geometry optimizations were carried out using a balanced polarized triple-zeta basis set (DEF2-TZVP). Exchange and correlations were described via the hybrid functional B3LYP \cite{becke, becke2, lyp}.
For the hyperfine calculations we switched to use a highly specialized double-zeta polarized basis set (EPR-II) \cite{barone}.
While we explain the hyperfine couplings in terms of the Fermi contact and dipolar contributions, our main focus will be on the eigenvalues of \Eqref{Atotal}, denoted as $A_{1}$, $A_{2}$ and $A_{3}$ with the general convention $|A_{1}| \geq |A_{2}| \geq |A_{3}|$.

We benchmark our \orca\ computations by calculating the HFIs for experimentally studied positive- and negative-ion anthracenes. These are characterized by spin doublets and are paradigmatic molecules akin to nanographenes. In \Tabref{tab:anthracene}, we provide the \orca-derived values for \carbon\ and \hydrogen\ isotropic HFIs for these molecules and compare the same with the experimental measurements \cite{bolton, weil_bolton}.
We use the HFIs of these molecules as a reference for our calculations.

We provide an alternative method to compute the hyperfine couplings based on parametrization technique using an \emph{empirical model} \cite{McCh.58.TheoryIsotropicHyperfine, KaFr.61.TheoreticalInterpretationCarbon13} in \Secref{sec:3-fitting-hfi}, we briefly mention a few important details pertaining to that method in this section. The crucial ingredient for the parametrization procedure requires the calculation of \pol\ at the carbon sites. Besides \orca\ we also use calculations based on \siesta\ DFT \cite{soler} and MFH descriptions \cite{dipc_hubbard} to compute the \pol\ for these molecules. 

Within \siesta\ calculations \cite{soler}, the site-resolved Mulliken \pol\ was obtained with a double-zeta-plus polarization (DZP) basis set and the PBE functional \cite{pbe}. We selected a 400 Ry cut-off for the real-space grid integrations and a 0.02 Ry energy shift as the cut-off radii for the generation of the basis set.

In case of the MFH computations, geometries were taken to be those from \textsc{Orca}, except for the large [$n$]triangulenes ($n>7$) where we took the hexagonal lattice with $d_\mathrm{C-C}=1.42$ \AA\ and $d_\mathrm{C-H}=1.10$ \AA. The \hubbard\ python package \cite{dipc_hubbard} was used to compute the \pol\ at the carbon sites by solving the MFH Hamiltonian \cite{rossier, Yazyev_2010, SaPaGi.22.SpinPolarizingElectron, OtFr.22.Carbonbasednanostructures} within a tight-binding description with single $\pi$ orbital and an onsite Coulomb repulsion parameter $U$
    \footnote{We point out that this method treats the carbon $\pi$-orbitals only, and, hence, the negligible polarizations on the H sites obtained with other methods are not computed.}.
The tight-binding models for the $sp^2$-carbon nanostructures were set up within the \textsc{Sisl} python package \cite{zerothi_sisl} with an orthogonal model parametrization such that the first-, second- and third-nearest-neighbor hopping matrix elements were $t_{1} = -2.7$ eV, $t_{2} = -0.2$ eV, $t_{3} = -0.18$ eV corresponding to interatomic distances below $d_{1} < 1.6$ \AA\ $<d_{2} < 2.6$ \AA\ $<d_{3} < 3.1$ \AA, respectively \cite{hancock2010}. 
The values for the empirical model reported in \Tabref{tab:anthracene} were based on MFH calculations with an onsite Coulomb repulsion parameter $U=4$ eV as explained in \Secref{sec:3-fitting-hfi}.

In the next section, within conventional \orcaxc, we derive the HFIs for several molecules belonging to the class of open-shell planar $sp^{2}$-carbon nanostructures and discuss numerous properties related to them.

\section{Hyperfine Interaction in open-shell planar carbon nanostructures}
\label{sec:HFInanographene}

In this section, our main aim is to give a description of the HFIs in several planar benzenoid as well as non-benzenoid hydrocarbons. We consider all the molecules to be fully hydrogenated with the plane of the molecule coinciding with the $xy$-plane. We use \orca\ to obtain the hyperfine matrix, for which we usually report its three eigenvalues containing contributions from both the isotropic Fermi contact and dipolar terms. In addition to this, we also provide a general picture of the site-resolved \pol\ in these molecules, since they are found to closely track the computed hyperfine coupling strengths.

\subsection{[2]triangulene}
\label{sec:hfi-2t}

\begin{figure}
    \centering
    \includegraphics[width=\columnwidth]{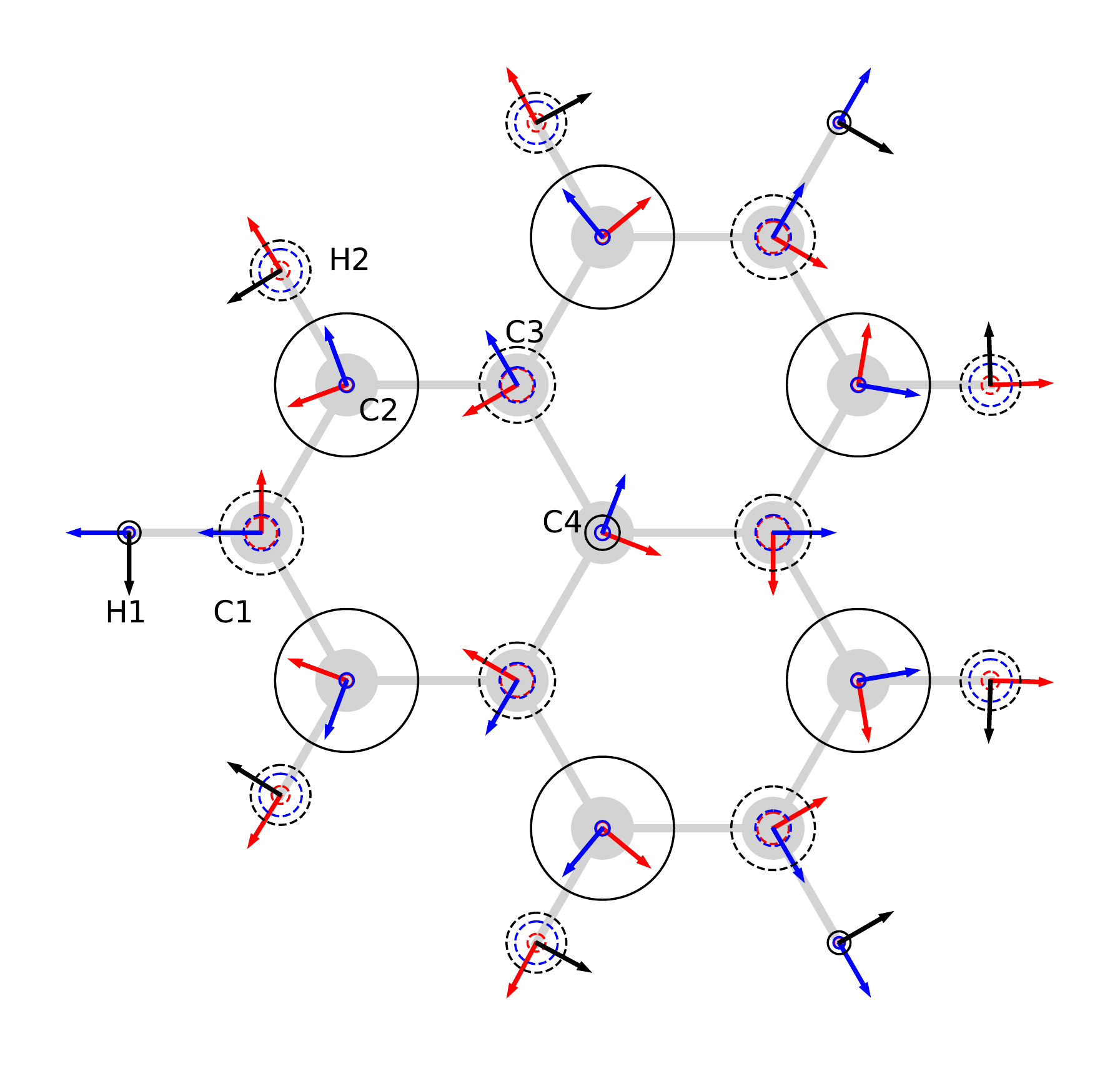}
    \caption{Depiction of the hyperfine tensors for all nuclei of [2]triangulene (2T, phenalenyl) oriented in the $xy$ plane, computed with \orcaxc. The radii of the circles scale with the absolute value of the three eigenvalues $|A_1|\geq|A_2|\geq|A_3|$ at that site, colored in decreasing order as black, blue, and red; solid (dashed) lines indicate positive (negative) eigenvalues.
    The arrows represent the orientation of the in-plane eigenvectors with their color matching the corresponding eigenvalue. The third eigenvector of $A$ always points in the (out-of-plane) $z$-direction.
    } 
    \label{fig:2Teig}
\end{figure}

We begin with a general description of the hyperfine matrix with regards to the size of the eigenvalues and the orientation of the eigenvectors. In \Figref{fig:2Teig}, we show our results for the case of a [2]triangulene (2T, also known as phenalenyl) molecule \cite{dataset}. 
The carbon atoms in this molecule form a bipartite structure (meaning that chemical bonds are only formed between atoms on opposite sublattices) and, further, there is an imbalance of one atom between the two sublattices \cite{rossier, Yazyev_2010, OtFr.22.Carbonbasednanostructures,turco}. Therefore, according to Lieb's theorem \cite{Li.89.TwotheoremsHubbard} for a half-filled $\pi$-electron system, this molecule is expected to display a doublet electronic ground state $S=(N_A-N_B)/2 = 1/2$ that couples to nuclear spins.

Several general features can be observed that also hold for the other molecules we have investigated. 
(1) For each nucleus, one of the three eigenvectors of the hyperfine tensor always points perpendicular to the plane of the molecule. For the carbon nuclei this out-of-plane eigenvalue is also the largest of the three. 
(2)
All three eigenvalues always have the same sign.
(3) There is a notable difference between the HFI of carbon nuclei belonging to the majority or minority sublattices of the molecule: for the former, where the unpaired electron spin is mainly localized, the three eigenvalues are positive, while they are negative for minority sites. The majority sites bonded to hydrogens show a very large anisotropy (very small in-plane eigenvalues: dipolar and isotropic contribution almost cancel each other), while for the minority sites, the three eigenvalues are closer in size, though still anisotropic.
(4) For the \hydrogen\ hyperfine tensors, their largest eigenvalue corresponds to an in-plane vector, perpendicular to the C-H $\sigma$ bond. The sign is opposite to that of the carbon site to which it is bonded. (5) The in-plane eigenvectors slightly deviate from the direction of the bond (to the attached \hydrogen) for the majority sites, showing that the electron spin density is not symmetric to that axis (see also \Appref{app:vec_2t}).

We summarize the magnitude of the three HFI eigenvalues and the corresponding site-resolved Mulliken \pol\ in \Tabref{tab:2t-hfi}. 
We note that due to the symmetries of the triangulene molecule and the $\pi$-spin density of the state we consider (reflection at the horizontal axis passing through H1 and C1 and \ang{120} rotations around C4), the HFI can be fully characterized by those of a small number of sites only. In case of 2T, the carbon sites labeled C1-C4 and the hydrogen sites H1 and H2 in \Figref{fig:2Teig} suffice.
The majority and minority sites with hydrogen atoms attached are represented by C2 and C1, respectively with the corresponding hydrogen indices being H2 and H1. The rest of the carbon atoms not associated with the hydrogens are represented by indices C3 (minority) and C4 (majority).
The hyperfine tensor at all other sites can be obtained by that of one of these six sites by applying the respective symmetry operation. In particular, the eigenvalues of hyperfine tensor at equivalent sites are the same, while the eigenvectors undergo the symmetry operations.

\begin{table}
\begin{tabular}{lrrrrr}
\toprule
$N$ & $A_{N,1}$ & $A_{N,2}$ & $A_{N, 3}$ & $A_N^\mathrm{iso}$ & $\Pi_N$ \\ 
\midrule
C1 & $-41.8$ & $-17.8$ & $-15.7$ & $-25.1$ & $-0.106$\\
C2 & 71.4 & 7.2 & 6.4 & $28.4$ &  0.247 \\
C3 & $-37.8$ & $-17.6$ & $-16.3$ & $-23.9$ & $-0.081$\\
C4  & 17.2 & 7.4 & 7.4 & $10.7$ & 0.044\\
H1 & 11.4 & 5.8 & 5.3 & $7.5$ & 0.000\\
H2 & $-29.9$ & $-21.2$ & $-8.9$ & $-20.0$ & 0.000 \\
\bottomrule
\end{tabular}
\caption{\label{tab:2t-hfi} The three hyperfine eigenvalues (in units of MHz) computed for \carbon\ and \hydrogen\ in [$2$]triangulene (2T, phenalenyl) with \orcaxc. The largest eigenvalue for \carbon\ is found for the out-of-plane direction, whereas for \hydrogen\ it is always in-plane. The indices refer to the positions given in \Figref{fig:2Teig}. The sign of the HFI for \carbon\ follows the sign of the Mulliken \pol\ ($\Pi_{N}$). However, for \hydrogen\ the sign of HFI is opposite to that of the carbon \pol\ to which it is bonded. }
\end{table}

Let us add a remark on the physical importance of the three eigenvalues. All three, together with the corresponding eigenvectors, are needed to fully specify the hyperfine tensor. We are particularly interested in the experimental situation of a single molecule exposed to a magnetic field of more than a few millitesla so that electron-nuclear spin exchange is suppressed. Then the HFI is determined by the matrix elements $A_{z'\alpha'}$: the $j$th nucleus contributes $\sum_\alpha A_{z'\alpha,j}I_j^{\alpha}$ to the Overhauser shift of the electron's Zeeman energy and is itself precessing in an effective field composed of the external magnetic field and the Knight field $K_j=S^{z'}(A_{z'x',j},A_{z'y',j},A_{z'z',j})$. For sub-tesla magnetic fields, $K_j$ dominates, the nucleus aligns with $K_j$ and the electron spin dephases ($T_2^*$) in a quasi-static Overhauser field as briefly discussed in \Secref{sec:application}. For 2T, and for all the molecules we discuss here, there is always one eigenvector in the $z$-direction (not necessarily the largest). This is due to the reflection symmetry of the spin density at the $z=0$-plane. 

In the next subsections, we report $A_{1}$ for all nuclei considered and also the two in-plane eigenvalues $A_2,A_3$ for \carbon. Since the orientation of the corresponding eigenvectors changes from nucleus to nucleus, an experiment with an in-plane magnetic field will for each nucleus see a hyperfine coupling strength in the interval between the two in-plane eigenvalues. For a perpendicular field, these eigenvalues determine, for example, the relaxation of the spins due to HFI and the electron-induced interaction between the nuclear spins that can be important for their long-term dynamics (and the decoherence of the electron spin).

\subsection{[\textit{n}]triangulenes}
\label{sec:ntriangulenes}

We now focus our attention to other molecules belonging to the class of [$n$]triangulenes (in short $n$T) that go beyond 2T. These fragments of graphene in the form of an equilateral triangle are characterized by $n$ hexagonal rings along their edges. One of the most interesting features of these molecules is their ground-state net spin $S = (n-1)/2$ growing with the molecular size \cite{rossier, wang, yazyev2008comment, hawrylak}. Following successful on-surface synthesis of these molecules \cite{niko,su2019atomically,mishra2019,Mishra2020,7tmishra, turco}, there has been great interest not only in the detection of the net spin, which gives rise to unconventional magnetism, but also in probing topological phases of matter in chains of triangulenes \cite{PhysRevLett.124.177201, Mishra2021, liu, frac_mishra}. 

\begin{figure}
    \centering
    \includegraphics[width=\columnwidth]{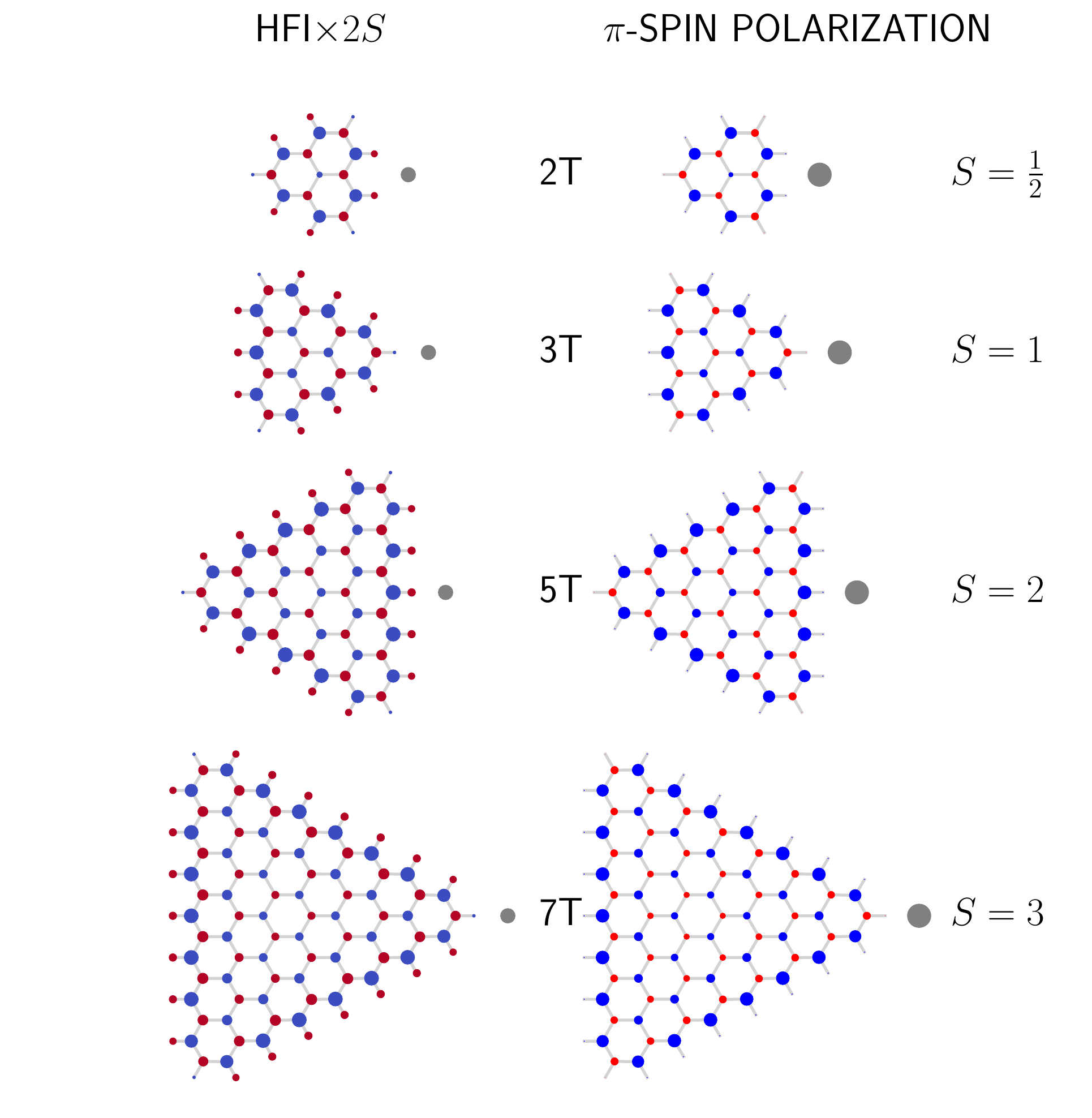}
    \vspace{-5mm}
    \caption{Maximum hyperfine eigenvalue (scaled by $2S$, left) and Mulliken \pol\ (right) for [$n$]triangulenes ($n$T) from \orcaxc. \hfispinpolplot } 
    \label{fig:hfi_spin_triangulene}
\end{figure}

Our main goal is to investigate the effects of the high-spin magnetic ground states on the HFIs. From Eqs.~(\ref{fermi})-(\ref{dipolar}), we see that the hyperfine couplings $A_N$ vary inversely with the number of unpaired electrons given by $2S$, thus this leads to a reduction in coupling strength for molecules with high-spin magnetic ground states. 
The highly symmetric nature of the molecules is also reflected in the appearance of identical HFI eigenvalues at the edges (\Figref{fig:hfi_spin_triangulene}). Out-of-plane HFIs for the \carbon\ nuclei for 2T to 7T corresponding to B3LYP computations are given in \Tabref{tab:13C_hfi}. The general trend shows higher values of maximal HFI for \carbon\ in case of 2T (spin doublet ground state) which then decreases with the increase in the number of unpaired electrons as we go from 3T (spin triplet) to 7T (septet ground state). 

The largest in-plane eigenvalues for \carbon\ nuclei appear at the minority sites that exhibit negative \pol\ (\Figref{fig:2Teig}). $A_{2}$, $A_{3}$ for 2T are $-17.8$ MHz and $-16.3$ MHz, respectively. For the rest of the triangulenes the modulus of these (always negative) eigenvalues for \carbon\ decreases with increasing $n$. Values of $|A_{2}|$ ($|A_{3}|$) are found to decrease from $12.0$ ($11.1$) MHz in 3T to $4.6$ ($4.2$) MHz in 7T within the B3LYP \cite{dataset}.

A similar pattern is also observed for the HFIs of the \hydrogen\ nuclei. As shown in \Figref{fig:2Teig}, the maximum HFIs for \hydrogen\ correspond to the in-plane vector perpendicular to the hydrogen's bond. We note that the larger contribution to the HFIs in this case is from the isotropic Fermi contact term. General values of \hydrogen\ HFIs for 2T are in the range of $A_{1} \in [-29.9, 11.4]$ MHz. Corresponding isotropic contributions are $A^{\mathrm{iso}} \in [-20, 7.5]$ MHz. 
For the other triangulene molecules, we find
$A_{1} \in [-18, 6.3]$ MHz \cite{dataset}.

\subsection{Topological end state in a graphene nanoribbon}
\label{sec:agnr}

Our next choice of molecule is the 7-carbon-atom wide armchair graphene nanoribbon (7AGNR) as a representative of the interesting family of graphene nanoribbons. These molecules are quasi-1D strips of graphene  with a wide variety of applications ranging from electronics to spintronics owing to their semiconducting behavior due to quantum confinement of charge carriers \cite{houtsma}. Atomically precise GNRs have well-defined edges and are successfully synthesized following on-surface bottom-up approaches \cite{cai2010atomically, talirz2016surface, zhou2020modified,de2018surface,chen2020graphene}. As in these works we consider the chemically stable case of a hydrogenated GNR, \ie, with hydrogen atoms attached to all edge carbons (ensuring no dangling $\sigma$-bonds). The 7AGNR is of particular interest because it possesses an electronic zero mode localized at each of the two (short) zigzag edges \cite{CaZhLo.17.TopologicalPhasesGraphene, OtFr.22.Carbonbasednanostructures} due to its nontrivial topology class.

\begin{figure*}
    \vspace{-5.5cm}
    \includegraphics[width=0.8\textwidth]{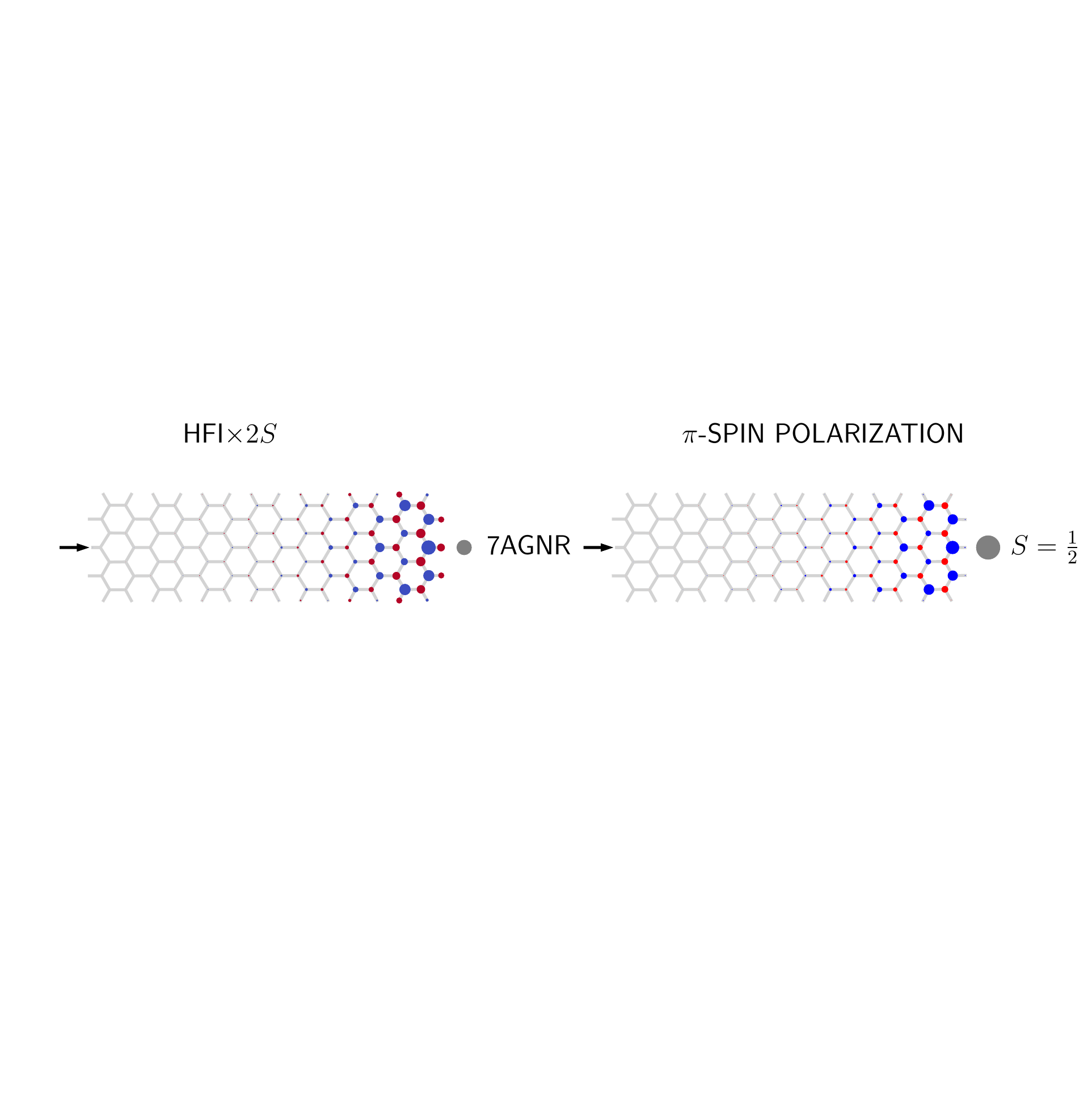}
    \vspace{-6.5cm}
    \caption{Maximum hyperfine eigenvalue (scaled by $2S$, left) and Mulliken \pol\ (right) for a topological end state in a 7AGNR from \orcaxc. The black arrow on the left side of the molecule indicates the $sp^3$-hybridized carbon site where an extra H has been attached (removing the topological end state at that terminus). \hfispinpolplot} 
    \label{fig:hfi_spin_AGNR}
\end{figure*}

In accordance with Lieb's theorem \cite{Li.89.TwotheoremsHubbard} for a sublattice-\emph{balanced} molecule, the ground state of this molecule represents a spin singlet corresponding to two half-filled topological end states located at the zigzag termini.
Here, however, we consider a situation with only one end-state by introducing a $sp^3$-hybridization of the central atom on one zigzag edge, which results in the effective removal of a carbon $\pi$-electron site and thus a doublet ground state with a single unpaired electron localized on the opposite edge (\Figref{fig:hfi_spin_AGNR}). (We briefly extend our considerations to the case of two end states in \Secref{sec:application}.) 
Quite naturally, the HFIs with maximum coupling appear on the edges and their magnitude is seen to decrease with increasing distance away from the edge of the AGNR. The out-of-plane HFIs for \carbon\ are given in \Tabref{tab:13C_hfi}. The largest in-plane eigenvalues for \carbon\ nuclei are $A_{2}\sim -19.2$ MHz and $A_{3} \sim -17.1$ MHz.
Similar to the \carbon, the HFIs for \hydrogen\ tend to be maximum on the edge and decrease with increasing distance from the edge of the AGNR. Appreciable values are in the range $A_{1} \in [-37.1, 8.7]$ MHz, with the dominant contribution once again coming from the isotropic Fermi contact interaction. The maximum HFIs for hydrogen are once again aligned in-plane (\Figref{fig:2Teig}).

\subsection{Non-benzenoid hydrocarbons}
\label{sec:indenes}

\begin{figure}
    \centering
    \includegraphics[width=\columnwidth]{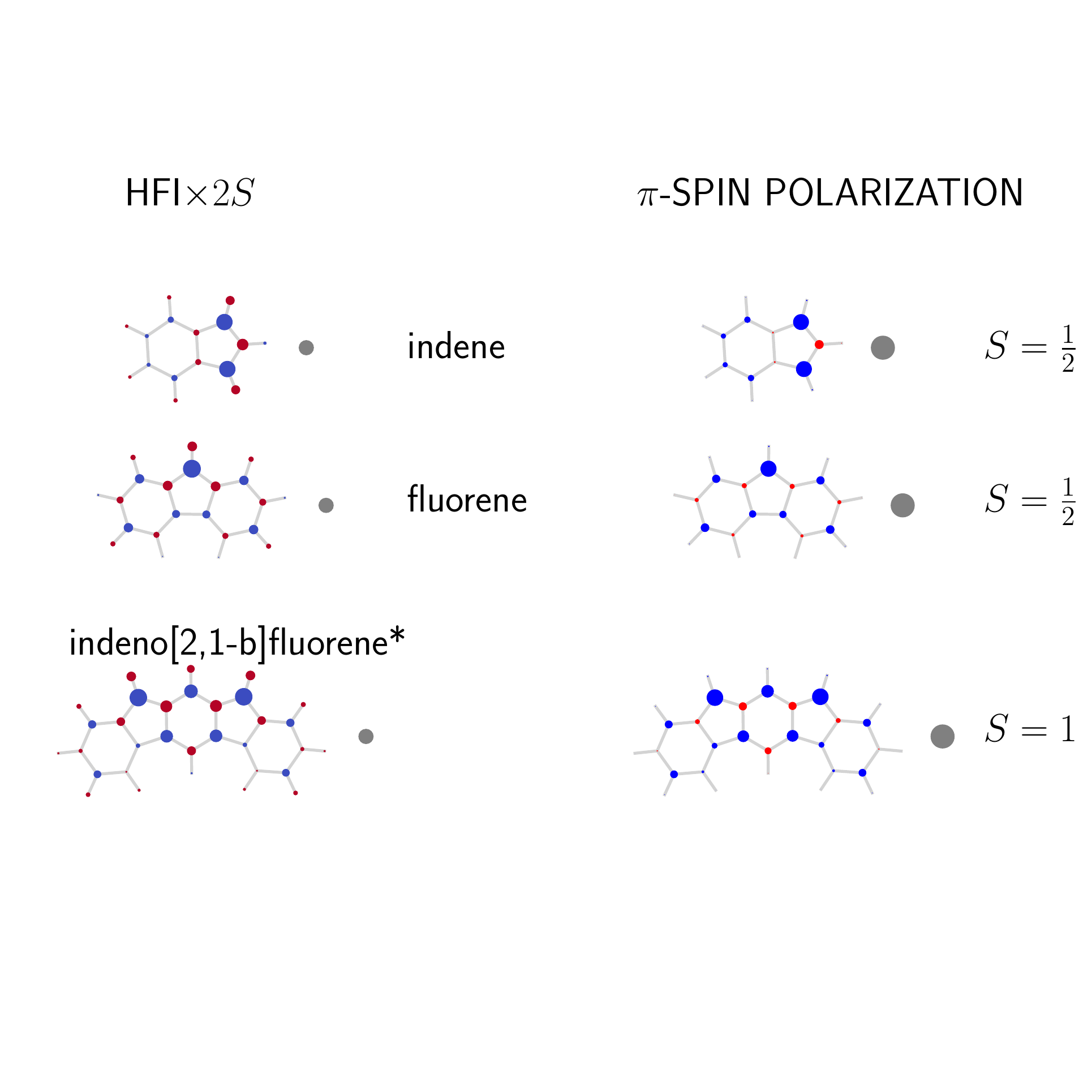}
    \vspace{-25mm}
    \caption{Maximum hyperfine eigenvalue (scaled by $2S$, left) and Mulliken \pol\ (right) for non-benzenoid hydrocarbons from \orcaxc. \hfispinpolplot} 
    \label{fig:hfi_spin_indeno}
\end{figure}

Finally, to go beyond the setting of benzenoid molecules considered so far, we provide the HFIs for a few molecules characterized by conjugated pentagon and hexagon rings (\Figref{fig:hfi_spin_indeno}), more specifically indene, fluorene, and indeno[2,1-b]fluorene* (the latter in the excited triplet state \cite{ShKiNa.13.Indeno21bfluorene20}, indicated here and in the following by a *-symbol).
While the molecules studied above are characterized by a bipartite structure (satisfying Lieb's theorem \cite{Li.89.TwotheoremsHubbard}), the molecules under consideration here break the bipartite character of the lattice owing to the presence of the pentagons at the edges \cite{ortiz16}. These non-benzenoid hydrocarbons have potential uses as active layers in electronic devices as well in molecular spintronics and non-linear optics \cite{fix, frederickson, nakano}.
Their successful on-surface chemical synthesis and characterization at the single-molecule level \cite{digio, mishra23} has led to a surge in research related to high-spin states, magnetism and topology \cite{mishra2022, ortiz}. These molecules can be seen as building blocks of chains of interacting localized (electron-)spin with interesting quantum correlations \cite{ortiz}.

The maximum HFI for \carbon\ is the out-of-plane eigenvalue ($A_1=A_{zz}$) and appears on the pentagons regardless of the specificity of the molecule. Amongst indene, fluorene and indeno[2,1-b]fluorene* the maximum HFI for \carbon\ is found for the indene and fluorene molecules which are characterized by spin doublets. 
The maximum value of $A_1$ of the \carbon\ HFIs are given in \Tabref{tab:13C_hfi}. The in-plane \carbon\ nuclei eigenvalues are largest for the indene and fluorene molecules, whereas it decreases for indeno[2,1-b]fluorene*. Within B3LYP the largest values for indene are $A_{2}\sim -26.1$ MHz and $A_{3}\sim-24.5$ MHz, whereas for fluorene it is $A_{2}\sim -24.8$ MHz and $A_{3}\sim -21.5$ MHz. For indeno[2,1-b]fluorene* using B3LYP we found $A_{2}\sim -17.2$ MHz and $A_{3}\sim -15.4$ MHz \cite{dataset}. 

The largest eigenvalues for \hydrogen\ are again related to in-plane eigenvectors and are highest in magnitude for the fluorene and indene with values $A_{1} \in [-61.7, 5.2$] MHz, $A_{1} \in [-54.0, 12.7]$ MHz, respectively. For indeno[2,1-b]fluorene* we find $A_{1} \in [-29.5, 4.6$] MHz  \cite{dataset}. This reduction compared to the other molecules in this group is justified because of the higher spin state of the indeno[2,1-b]fluorene*, compared to indene and fluorene.

\subsection{Trends across the studied nanostructures}
\label{sec:sum_hfi}

\begin{table}
\begin{tabular}{lcrrrr}
\toprule
Molecule & $S$ & $A_{N,1}$ & $A_N^\mathrm{iso}$ & $\eta/2S$ & $\Pi_{N}$ \\ 
&& $\times 2S$ & $\times 2S$\\ 
\midrule
2T & 1/2 & $71.4$ & $28.4$ & 2.36 & 0.247  \\
3T & 1 & $88.5$ & $33.4$ & 2.38 &0.314 \\
4T & 3/2 & $93.7$ & $35.4$ & 2.33 & 0.329\\
5T & 2 & $96.3$ & $36.5$ & 2.33 & 0.336\\
6T & 5/2 & $95.0$ & $36.0$ & 2.35 & 0.331 \\
7T & 3 & $94.2$ &$35.7$ & 2.36 & 0.329\\
7AGNR$\dagger$ & 1/2 & $88.8$ & $32.4$ & 3.19 &0.325 \\
indene & 1/2 & $115.4$ & $41.2$ & 2.40 &0.436 \\
fluorene & 1/2 & $137.6$ & $49.2$ & 3.51 & 0.577\\
\multicolumn{1}{p{20mm}}{indeno[2,1-b]-fluorene*} & 1 & $132.2$ & $47.7$ & 2.75 & 0.472\\
\bottomrule
\end{tabular}
\flushleft $\dagger$ topological end state;  *excited triplet state
\caption{\label{tab:13C_hfi} \carbon\ largest HFI (scaled by $2S$, in units of MHz) appearing at specific sites $N$ in $sp^2$-carbon nanostructures computed within \orcaxc. For planar molecules oriented along $xy$ plane, the largest HFIs are found perpendicular to the plane of the molecule, \ie, $A_{N,1}=A_{N,zz}$.} 
\end{table}

\begin{figure}
    \centering
    \includegraphics[width=\columnwidth]{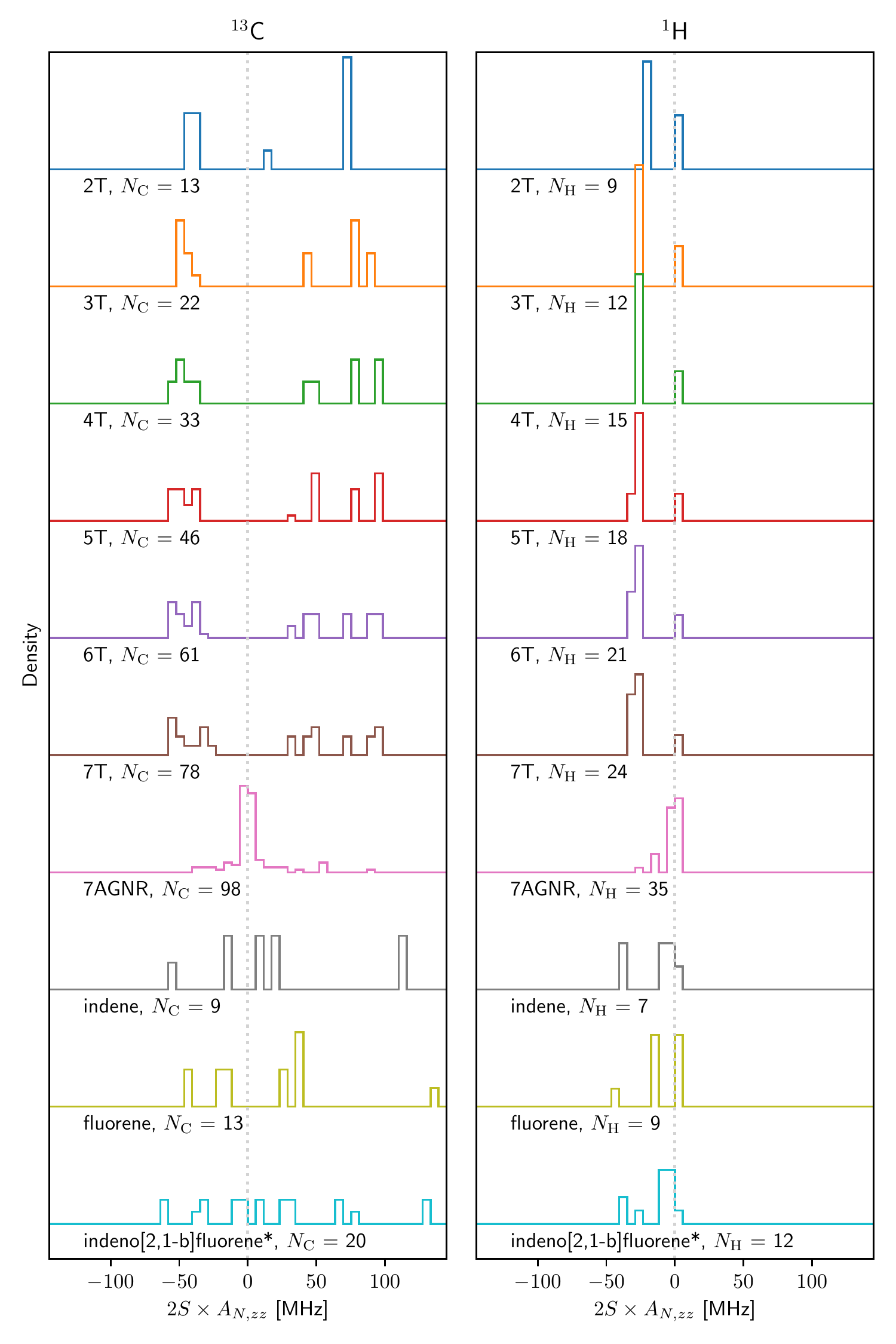}
    \caption{Normalized histogram plots of $2S\times A_{zz}$ (bin size 5.8 MHz) for \carbon\ (left) and \hydrogen\ (right) from \orcaxc. $N_\mathrm{C}$ and $N_\mathrm{H}$ represent the number of C and N atoms in the molecules, respectively.}
    \label{fig:histplot_Azz}
\end{figure}

So far, we discussed the HFI for these molecules, which is given by real-space integrals over the electron spin density. But an important role is played by the \pol. In Figs.~\ref{fig:hfi_spin_triangulene},~\ref{fig:hfi_spin_AGNR},~\ref{fig:hfi_spin_indeno} in addition to the HFIs we also show the atom-resolved \pol, cf.~\Eqref{mullikenpi}, which are computed within Mulliken spin-population analysis in \orca\ \cite{dataset}. The red and the blue blobs depict the negative and positive \pol\ (same for HFIs), respectively. 
As we see the HFIs follow the \pol\ in terms of the location of the maximum HFI (out-of-plane $A_{zz}$) as well the sign and relative size. The maximum HFI for these molecules is associated with the maximum atom-resolved \pol\ $\Pi_{N}$ at a specific site. 
For \carbon\ this trend can be seen from \Tabref{tab:13C_hfi} with $2SA^{\mathrm{iso}}_N/\Pi_N=100\pm15$ MHz. Although the Fermi contact contribution $A^{(\mathrm{iso})}$ is not recoverable within real-space integrals using only $\pi$-electrons, we do observe that it is also related to the \pol. We will explore this relation between HFI and \pol\ explicitly in \Secref{sec:3-fitting-hfi} below.

Another interesting feature about these molecules is the \textit{Inverse Participation ratio (IPR)}. It is defined via the Mulliken $\pi$-spin polarization $\Pi_j$ as
\begin{equation}
    \frac{1}{\eta} = \sum_{j}\bigg(\frac{\Pi_{j}}{2S}\bigg)^{2},
\end{equation}
where we divide by $2S$ to get the contribution per electron spin. The quantity $\eta$ can then interpreted as the effective number of participating sites over which an electron spin is distributed. We observe that for the class of molecules discussed, this factor $\eta \approx 3$, indicating that the unpaired electron spin is occupying around three sites.
The maximum HFI for the benzenoid molecules remains constant for such degree of localization. For the non-benzenoid molecules, we observe a higher HFI for a constant magnitude of localization of the electron spin which could be related to the larger magnitudes of \pol\ (as seen in \Tabref{tab:13C_hfi}).

Some further general properties of the calculated HFI can be observed: the hyperfine tensor has a common sign (the three eigenvalues are either all positive or all negative) for all nuclei we considered. This implies that the isotropic part of the HFI is at least larger than $\|A_j^{\mathrm{dip}}\|/2$ in all cases. For most nuclei, the dipolar part is significant, \ie, $A_1\not\approx A_2,A_3$; specifically, we find the ratio $(A_2+A_3)/(2A_1)$, which ranges from $-1/2$ for fully dipolar to $+1$ for fully isotropic coupling, to lie in a range $[0.01, 0.94]$ very similar to that found for 2T (cf.~\Tabref{tab:2t-hfi}). The sign of $A$ is the same as the sign of the \pol\ associated to that site for \carbon\ nuclei (with the exception of the two most weakly coupled carbon sites in indenofluorene, cf.~\Figref{fig:hfi_spin_indeno}). Additionally, we mention that there are small anomalies such as deviations from neighboring sites having alternating sign of the HF tensor in the non-benzenoid structures. These molecules differ from the others considered here also in that they do not have well-defined majority and minority sites given that the carbon lattice is not bipartite.

The anisotropy of the HFI could be exploited in different ways. For example, the Overhauser field variance (and the $T_2^*$ dephasing time induced by quasistatic nuclear spin bath) will depend on the orientation of the external magnetic field, allowing the optimization of $T_2^*$ (cf.~\Secref{sec:application}). More precisely, this can be seen as a trade-off between dephasing and relaxation, since orienting the magnetic field to reduce the HFI parallel to it will increase the transversal terms that enable electron-nuclear spin flips.

\Figref{fig:histplot_Azz} gives a view of the distribution of all the $A_{zz}$ eigenvalues in our molecules which contribute to the Overhauser field in an external field along $z$. Due to the symmetry of the considered molecules and electronic states, there are many identically coupled nuclei, leading to isolated peaks in the histograms, especially for \hydrogen-nuclei. That has consequences for the expected time-evolution of electron spin coherences: instead of simple dephasing due to the nuclear spin bath, one would expect collapses and revivals of the coherence with the few characteristic frequencies (sums of certain integer multiples of the $A_{zz,j}$) of the spin bath. Note that for a magnetic field with an in-plane component, this exact symmetry is lost due to the different orientation of the hyperfine tensors of equivalent nuclei. 

For larger molecules, and especially for the \carbon-nuclei, a much broader distribution around 0 is found, which limits or completely suppresses the appearance of revivals \cite{coish, Schliemann03}. Note that one can consider three distinct realizations of the molecules we consider here: if the natural abundance of \carbon\ is considered, only 1\% of the carbon sites have a nuclear spin and the hyperfine dynamics and spectrum will differ significantly depending on whether a strongly coupled carbon site has a nuclear spin or not and the left column in \Figref{fig:histplot_Azz} gives the probability with which certain carbon HFI can be expected. The smallest HFI is encountered in purely-$^{12}$C molecules, in which we only have to deal with the hydrogen spin bath (right column of \Figref{fig:histplot_Azz}). To maximize HFI one could work with purely-\carbon\ molecules (combining both columns of \Figref{fig:histplot_Azz}). 

We devote the next section to describing a quantitative and qualitative relation between the HFIs and the \pol.

\section{Empirical Parametrization of HFI from \pol}
\label{sec:3-fitting-hfi}
In this section, we discuss a parametrization procedure that relates the HFIs to the atom-resolved \pol.
As remarked above, the hyperfine tensor is defined via a spatial integral over the full electron spin density given by Eqs.~(\ref{fermi})-(\ref{dipolar}). As shown by Karplus and Fraenkel \cite{KaFr.61.TheoreticalInterpretationCarbon13}, a good quantitative understanding of $^{13}$C in $\pi$-radicals can be obtained from a few-parameter fit taking into account only the $\pi$-spin polarization on the carbon atom at the site of the nucleus and its nearest neighbors. This approach has found applications in various contexts, including endohedral fullerenes \cite{SeBaDu.98.Electronspinresonance}, adatoms on graphene  \cite{LeFoAy.03.MagneticPropertiesDiffusion}, and nanographenes \cite{Ya.08.HyperfineInteractionsGraphene}. We apply and extend this fitting procedure in the following ways: (i) confirm it for molecules where it had not been tested before; (ii) extend it by also fitting the dipolar contribution; (iii) extend it to different methods for obtaining the carbon \pol.
This may be interpreted as a consequence of a linear relation between the total spin density at $\mathbf{r}$ and the $\pi$ contribution (at $\mathbf{r}$ and neighboring locations).

\subsection{General fitting procedure}
For the set of molecules ($\mu=1,\dots,M$) presented in \Secref{sec:HFInanographene}, we take the hyperfine tensors from \orca\ of the $N^{(\mu)}$ nuclear spins for each molecule as reference \cite{dataset}.
To obtain the fit parameters we then compute for each molecule the carbon Mulliken \pol\ associated to all the $N^{(\mu)}$ sites using different methods (\orca, \siesta, MFH; computational details for carbon \pol\ is given in \Secref{sec:2.3-procedure}), which yields a population vector $\Pi^{(\mu)}$ with $N^{(\mu)}$ components.
For each species $\alpha=({}^1\mathrm{H}, ^{13}\mathrm{C})$, of which there are $N^{(\mu,\alpha)}$ nuclei in the molecule $\mu$, we use four parameters ($F_0^{(\mu,\alpha)},F_1^{(\mu,\alpha)}$, $D_0^{(\mu,\alpha)},D_1^{(\mu,\alpha)}$) to fit the hyperfine tensor.  This is done by obtaining separately (for each method of obtaining carbon $\Pi^{(\mu)}$, for each molecule $\mu$, for each nuclear species $\alpha$, and for the isotropic and anisotropic part of $A$) the least-mean-square fit to the hyperfine tensor as 
computed by \orca. \emph{E.g.}, for \carbon\ nuclei, we have
\begin{widetext}
\begin{align}\label{eq:fitparams}
(F_0^{(\mu,\mathrm{C})},F_1^{(\mu,\mathrm{C})})
    &=\mathrm{argmin}_{f_0,f_1}\left\{\sum_{j=1}^{N^{(\mu,\mathrm{C})}}
       \left|A^{\mathrm{iso},\mu}_j-\frac{1}{2S}\left(f_0\Pi^{(\mu)}_j
        +f_1\sum_{l\in D_R(j)}\Pi^{(\mu)}_l\right)\right|^2\right\},\\
(D_0^{(\mu,\mathrm{C})},D_1^{(\mu,\mathrm{C})})
    &=\mathrm{argmin}_{d_0,d_1}\left\{\sum_{j=1}^{N^{(\mu,\mathrm{C})}}
        \left\|A^{\mathrm{dip},\mu}_j-\frac{1}{2S}\left(d_0Q_0\Pi_j^{(\mu)}
        +d_1\sum_{M\in D_R(j)}Q_{\mathbf{v}_M}\Pi_M^{(\mu)}\right)\right\|^2\right\},
\end{align}
\end{widetext}
where $Q_0=\mathrm{diag}(-1,-1,2)/2$ and $Q_{\hat{\mathbf{v}}_M}=(3\hat{\textbf{v}}_{M}^{T}\hat{\textbf{v}}_{M} -   \hat{v}_{M}^{2}\mathbbm{1})/\hat{v}_{M}^{5}$ are the dipolar matrices with $\hat{\textbf{v}}$ giving the vector from site $j$ to site $M$ in units of Bohr such that the matrix is dimensionless. $D_R(j)$ denotes the set of sites within a radius $R$ around $j$. We work with $R=1.7$ \AA, which would include the three closest sites on a standard graphene lattice.\footnote{As we mentioned above, our fitting procedure aims to match the dipolar term by taking into account a term proportional to the contributions from the nearest-neighbor $\pi$-spin densities only. For the mean-square error, this turned out to be a better choice than including more neighbors. However, in some cases, this choice will miss certain small details such as the orientation of the in-plane eigenvectors in the 2T molecule, see \Appref{app:vec_2t}.}
To account for the negligibly small \pol\ at the hydrogen sites, for \hydrogen\ nuclei the $f_0$ term is zero and $d_0$ multiplies $\Pi^{(\mu)}_{n(j)}$ the \pol\ associated with the nearest carbon site instead of $\Pi^{(\mu)}_j$. 
This gives a different set of seven fit parameters for each molecule and for each method (we do not include a subscript labeling the method to not overload notation).

\begin{figure}
    \centering
    \includegraphics[width=\columnwidth]{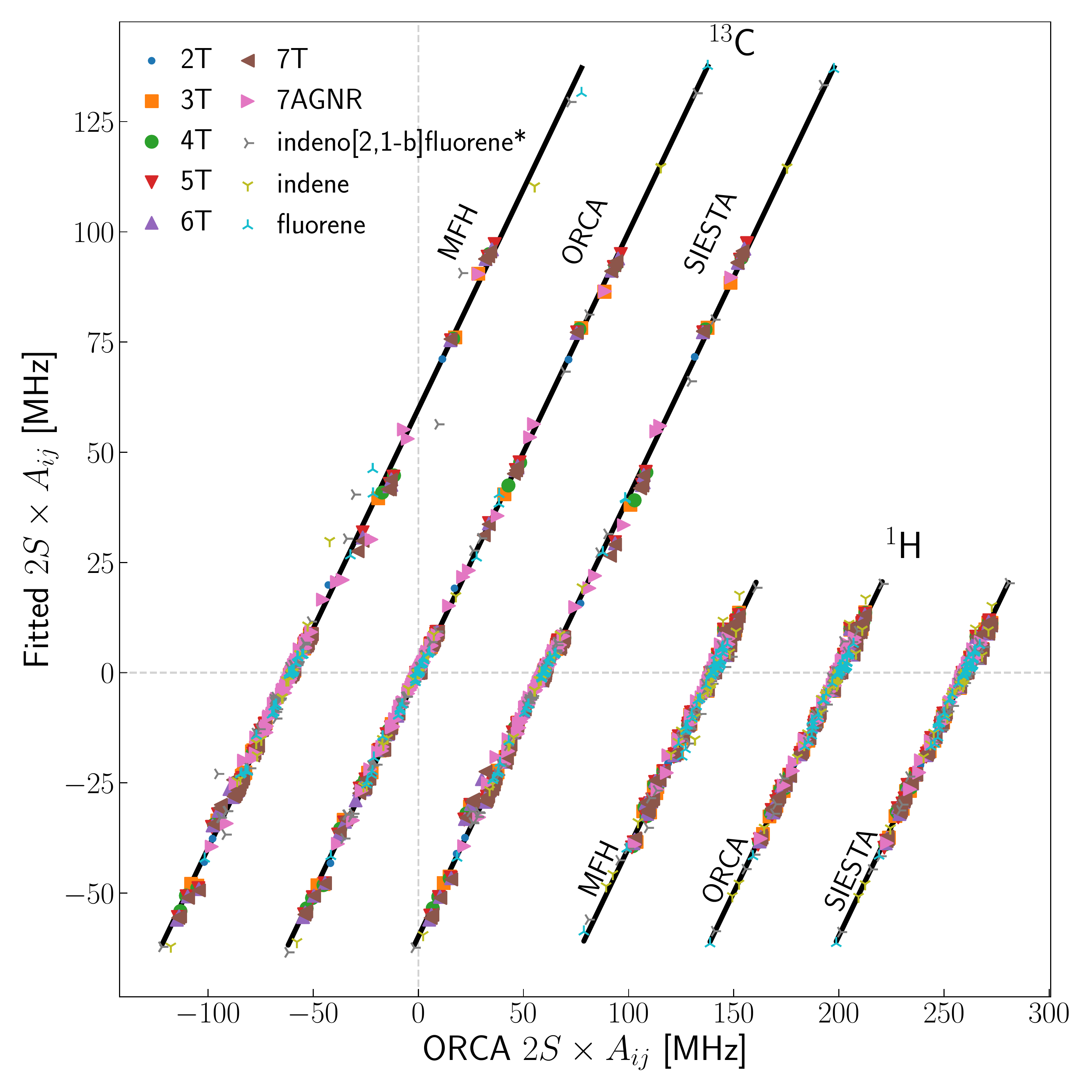}
    \caption{Fitted \textit{vs} \orca-calculated HFI matrix elements $A_{ij}$ within B3LYP.
    For clarity, the calculations are horizontally offset with respect to the \carbon-\orca\ fit.
    The black lines are linear fits.} 
    \label{fig:fit}
\end{figure}

As displayed in \Figref{fig:fit}, the results all lie within a few percent of each other. This suggests that it is possible to obtain a set of \textit{generic $sp^2$-hydrocarbon fit parameters} that should work well for all such molecules. For each method, we average the parameters obtained for the different molecules (for each species $\alpha$ of nucleus:
$F^{(\alpha)}_u=\mathrm{average}_{\mu}\{F_u^{(\mu,\alpha)}\}$ and same for $D^{(\alpha)}_u$) as shown in \Tabref{tab:fit_params_C}, giving rise to the general formulas for \carbon\ nuclei \cite{KaFr.61.TheoreticalInterpretationCarbon13}
\begin{align}\label{model:fermi}
A^\mathrm{iso}_N 
    &= \frac{1}{2S}\bigg[F_{0}^{(\mathrm{C})} \Pi_N 
        + F_{1}^{(\mathrm{C})} \sum_{M\in D_R(N)} \Pi_M\bigg],\\
\label{model:dipolar}
A^{\mathrm{dip}}_N 
    &=\frac{1}{2S}\bigg[D_0^{(\mathrm{C})} Q_0\Pi_N
        +D_1^{(\mathrm{C})}\sum_{M\in D_R(N)}Q_{\mathbf{v}_M}\Pi_M\bigg],
\end{align}
where we have suppressed the superscript labeling the molecule in the populations. For \hydrogen, $F_0^{(\mathrm{H})}=0$ \cite{McCh.58.TheoryIsotropicHyperfine} and the dipolar equation changes in that $Q_0$ is multiplied by the Mulliken \pol\ referring to the nearest-neighbor \carbon\ site. 

\begin{table*}
\begin{tabular}{lrrrrrrrrrrrrrr}
\toprule
& \multicolumn{3}{c}{\carbon\ Fermi contact} & \multicolumn{3}{c}{\carbon\ dipolar}  
& \multicolumn{3}{c}{\hydrogen\ Fermi contact} & \multicolumn{3}{c}{\hydrogen\ dipolar}\\
\cmidrule(r){2-4} \cmidrule(l){5-7} \cmidrule(l){8-10} \cmidrule(l){11-13}
Method \hspace{2cm} & $F_{0}^{(\mathrm{C})}$ & $F_{1}^{(\mathrm{C})}$ & RMSE & $D_{0}^{(\mathrm{C})}$ & $D_{1}^{(\mathrm{C})}$ & RMSE 
&& $F_{1}^{(\mathrm{H})}$ & RMSE &  $D_{0}^{(\mathrm{H})}$ & $D_{1}^{(\mathrm{H})}$ & RMSE \\
\midrule
\orca\ (B3LYP) &  89.2 & $-33.4$ & 0.3 & 179.7 &  1.7 & 1.6 & \hspace{7mm} & $-81.9$ & 0.3 & 23.6 & 36.8 & 2.2\\
MFH ($U=4$ eV) &  86.7 & $-33.1$ & 0.8 & 173.3 &  4.6 & 2.1 &              & $-78.7$ & 0.6 & 22.6 & 35.3 &  2.6  \\
MFH ($U=3$ eV) & 104.9 & $-42.3$ & 1.0 & 196.5 & 21.9 & 3.4 &              & $-89.7$ & 0.9 & 25.4 & 40.5 & 2.3 \\
\siesta\ (PBE) & 103.9 & $-40.5$ & 0.4 & 202.0 & 12.8 & 1.3 &              & $-92.5$ & 0.3 & 26.4 & 43.9 & 2.1 \\
\bottomrule
\end{tabular}
\caption{\label{tab:fit_params_C} Generic $sp^2$-hydrocarbon fit parameters for \carbon\ and \hydrogen\ HFIs (in units of MHz). $F_{0}^{(\alpha)}$, $F_{1}^{(\alpha)}$ describe the isotropic Fermi contact contribution, whereas $D_{0}^{(\alpha)}$ and $D_{1}^{(\alpha)}$ characterize the anisotropic hyperfine coupling matrices. Root-mean-square errors (RMSE) for the Fermi contact and dipolar contributions to HFI are also provided.}
\end{table*}

We characterize the quality of the fit by the root-mean-square-error (RMSE) with respect to the \orca\ results, which  for \carbon\ (\hydrogen) is  0.3 (0.3) MHz for the isotropic part and 1.6 (2.2) MHz for the dipolar. More details on the RMSE for individual methods is given in \Tabref{tab:fit_params_C}. Across all the reference
molecules, we find a maximal difference between the fit and the
reference value matrix elements of 5.1 (4.2) MHz for dipolar
interaction of \carbon\ (\hydrogen). 
These extreme values are obtained for the MFH fit in indene. For fits based on \siesta\ and \orca\ spin densities, this
maximum error is reduced to 1.0 (3.8) MHz and 1.4 (4.0) MHz, respectively.

The \pol\ from \Eqref{mullikenpi} is what we need for the fit. We note that \orca\ (B3LYP) as well as MFH ($U=4$ eV) generate significantly larger spin densities as compared to \siesta\ using the PBE functional or to MFH ($U=3$ eV). This is reflected in the $\sim10-25\%$ larger fit parameters obtained for the latter two. 

The fit parameters and associated error estimates that we report are based on the 393 \carbon\ and 162 \hydrogen\ nuclei in the molecules discussed in \Secref{sec:HFInanographene}. In case of the dipolar hyperfine tensor there are four independent matrix elements, so our fit aims to match in total 1572 (648) independent matrix elements. In \Tabref{tab:fit_params_C} we report the fit parameters obtained for the different methods as well as the associated errors (with respect to the \orca-calculated reference values). The fit optimizes the root-mean-square error per molecule and the table gives the average of this number taken over the ten molecules considered. The main point of these numbers is to show that the fitted numbers are typically much closer to the reference values than the latter is to the few experimentally available data points (in the acenes, cf.~\Secref{sec:2.3-procedure}). This suggests that with the use of fit parameters rather simple electronic calculation like MFH can estimate HFI and are thus comparable in quality to those obtained from more sophisticated (and computationally expensive) ones. As a first confirmation, we apply the MFH-derived fit parameters to the anthracene ions (using $U=4$ eV) as reported in (\Tabref{tab:anthracene}) and find it matching with the calculated \orca\ (EPR-II/B3LYP) values within a rms error of 1.00 MHz for the positive and 1.61 MHz for the negative ion, respectively.

We note that for all methods (i) the isotropic part is matched better than the dipolar one (maybe not surprising given that the former fits a number only, while the latter fits a matrix with four independent parameters), (ii) small molecules show larger deviations than large ones (which may be related to the reduced hyperfine coupling in larger structures, since small hyperfine tensors lead to small rms error), (iii) strong localization of the electron (as in the case of the AGNR and the small molecules) leads to larger error (since it entails a large hyperfine tensor that can contribute disproportionally to rmse) and (iv) the non-benzenoid structures show larger fitting error, especially for the MFH (which may be related to the deviation from the graphene structure for which the $\pi$-MFH model is best suited). 
Moreover, we learn that all three methods match the reference values well. This indicates that the parametrization technique, and thus the underlying \emph{empirical model} \cite{KaFr.61.TheoreticalInterpretationCarbon13, McCh.58.TheoryIsotropicHyperfine} given by Eqs.~(\ref{model:fermi})-(\ref{model:dipolar}), is well-suited to study HFI in these classes of benzenoid and non-benzenoid carbon nanostructures.

We must mention that while all the electronic structure methods have produced comparable results, there is an advantage of using the MFH approach. This method is simpler (owing to the presence of only a single $\pi$ orbital) and is thus computationally inexpensive and can be applied even to large systems which cannot be afforded within \orca. In \Secref{sec:large-molecules-MFH}, we will use the fit parameters derived within \hubbard\ (MFH $U = 4$ eV) to explore the effect of system size on HFI in $[n]$triangulenes for $n<40$, \ie, for molecules containing up to 1798 atoms.

\subsection{Physical interpretation \& intuitive understanding of HFI}
Let us add a brief discussion on the physical interpretation of the fitting formulas Eqs.~(\ref{model:fermi})-(\ref{model:dipolar}) by relating the HFI predicted by the fitting formula to the HFI that would be caused by a net spin polarization in an atomic orbital of the nucleus considered. The prefactor $ \mathcal{P} = 4/3\mu_0\gamma_{e} \gamma_{N}$ in \Eqref{fermi} takes the value $\mathcal{P}^{(\mathrm{H})} = 8940$ MHz-Bohr$^{3}$ for \hydrogen. A fully spin-polarized $1s$ orbital (probability density $1/\pi$ Bohr$^{-3}$ at the nucleus) gives rise to $A_0^{\mathrm{iso,H}}=\mathcal{P}^{(\mathrm{H})}\rho(\bR_N)=2846$ MHz. Similarly, $\mathcal{P}^{(\mathrm{C})} = 2248$ MHz-Bohr$^{3}$ and considering an $2s$ electron for an effective charge $Z_{\mathrm{eff}}\approx3.2$, we get $A_0^{\mathrm{iso, C}}=2964$ MHz. Expressing $F_0^{(\mathrm{H})}$ and $F_0^{(\mathrm{C})}$ in these units, respectively, we see that a given (MFH $U=4$ eV-calculated) Mulliken \pol\ $\Pi_N$  associated to carbon site $N$ corresponds (in terms of the induced HFI) to a induced $2s$-spin polarization of $F^{(\mathrm{C})}_0/A_0^{\mathrm{iso,C}}\approx0.029 \Pi_N$ at this site and to a $1s$-spin polarization in an attached hydrogen of $F^{(\mathrm{H})}_0/A_0^{\mathrm{iso, H}}\approx-0.028 \Pi_N$. And the $F^{(\mathrm{C})}_1$ terms imply that a $\Pi_M$ on nearest-neighbor sites to a carbon site $N$ gives rise to the contact HFI corresponding to a $-0.011\Pi_M$-polarized $2p_z$ orbital.

One can make a similar comparison for the dipolar fit with the HFI induced by a polarized $2p_z$ electron (for H or C). 
The corresponding dipolar HFI would be proportional to $Q_0$ and of strength $A^{\mathrm{dip},0}_\mathrm{H}=\mathcal{P}^{(\mathrm{H})}/(320\pi)$ 
and $A^{\mathrm{dip},0}_\mathrm{C}\approx \mathcal{P}^{(\mathrm{C})} Z_{\mathrm{eff}}^3/(320\pi)$, respectively 
    \footnote{The factor $Z^3/120$ in the expressions for $A^{\mathrm{dip},0}_\mathrm{H},A^{\mathrm{dip},0}_\mathrm{C}$ is the value of the spatial integral in \Eqref{dipolar} for the $2p_z$ hydrogenic spin density (for effective charge $Z=1, Z=3.21$, respectively).},
so the values of the fitting parameters $D_0^{(\mathrm{H})}, D_0^{(\mathrm{C})}$ imply that a Mulliken \pol\ $\Pi_N$ at a carbon site corresponds to an effective $2.5\Pi_N$ polarization (of a $2p_z$ orbital with $Z_\mathrm{eff}=3.2$) and a $2.3\Pi_N$ $2p_z$-polarization of an adjacent \hydrogen.
The final term in \Eqref{model:dipolar} means that all the carbon Mulliken \pol\ at sites $j\in D_R(N)$ contribute to HFI of a nucleus at site $N$ like a point dipole at site $j$ would.
The fit parameters then imply that $\Pi_j$ leads to the dipolar HFI with a nucleus at $N$ that would be caused by a point dipole of strength $0.1\%\Pi_j$ ($3.6\%\Pi_j$) at $j$.

The empirical model allows to give simple explanations for the size and shape of the hyperfine tensors. To see this, let us consider again the [2]triangulene depicted in \Figref{fig:2Teig}. For the hydrogen nuclei, only the Mulliken \pol\ $\Pi_j$ of the carbon site to which it is attached is important for the HFI, which according to our fit is given by three terms, all proportional to $\Pi_j$. The sign of $A$ is determined by the isotropic term and hence is negative for \hydrogen\ attached to majority sites. Since $F_1^{(\mathrm{H})}$ and $D_0^{(\mathrm{H})}$ have opposite sign, the in-plane eigenvalues are enhanced while $A_{zz}$ is reduced. The in-plane symmetry is then broken by the $D_1^{(\mathrm{H})}$-term, whose positive eigenvalue points along the bond to the hydrogen and thus reduces the corresponding eigenvalue of $A$, making the orthogonal one the largest of the three.

For the \carbon\ nuclei, both the on-site \pol\ $\Pi_N$ and those of the nearest-neighbor carbon sites $j\in nn(N)$ are important. Again, the sign of $A$ is determined by the isotropic term $F^{(\mathrm{C})}_0\Pi_N+F^{(\mathrm{C})}_1\sum_{j\in nn(N)}\Pi_j$, which is positive for $N$ being a majority site and negative otherwise. The marked difference in the anisotropy of $A$ (majority sites are highly anisotropic, minority sites almost isotropic) follows from how the on-site dipolar term $D^{(\mathrm{C})}_0$ contributes: for majority sites $\Pi_N$ is large and as $D_0$ is large and positive it strongly increases $A_{zz}$ and strongly reduces the in-plane eigenvalues. For minority sites, on the other hand, $\Pi_N$ is small and the degeneracy of the in-plane eigenvalues is broken by the anisotropic distribution of nearest-neighbor carbon \pol. This effect is small (since $D_1$ is small), and most pronounced for minority sites (which have larger $\Pi_{j\in nn(N)}$). For those, the two strongly polarized neighboring majority sites (which are equivalent and therefore have equal $\Pi_j$ and contribute equally to $A_N$) add up to a dipolar contribution that enhances the eigenvalue in the direction of the third neighbor. For the majority sites, the neighboring sites are no longer equivalent and no simple alignment of the eigenvectors with the lattice directions is found. This shows that one can obtain reasonably good insight into the expected hyperfine tensors by reasoning directly from the carbon \pol\ and geometry of the molecule.

\section{Extension to large molecules: Application of Fitting Procedure}
\label{sec:large-molecules-MFH}
One of the expected advantages of the fitting procedure is that it allows to make predictions for hyperfine interactions in large structures for which the effort to do a full \orca\ calculation cannot be afforded.
In this section, we will showcase the use of the fit parameters extracted in the previous section to study the scaling behaviour of the hyperfine interaction for [$n$]triangulenes for large $n$.

We use the fit parameters found in \Tabref{tab:fit_params_C} from MFH ($U = 4$ eV) to find HFI for [$n$]triangulenes with $n>7$ up to $n=39$, \ie, a molecule with 1798 atoms ($N_\mathrm{C}=1678$ and $N_\mathrm{H}=120$).
To illustrate the computational speedup, the MFH result for 39T can be obtained on a standard laptop within a few minutes while \orca\ for 7T with 102 atoms ($N_\mathrm{C}=78$ and $N_\mathrm{H}=24$) takes several hours with 16 cores on a supercomputer.

Before we turn to the results of the computation, consider what can be expected: there are $n-1$  unpaired electron spins in [$n$]triangulene giving rise to its spin $S=(n-1)/2$ ground state. Thus as we go from $n$ to $n+1$, one unpaired spin is added. In the \textit{spin-chain model} we can imagine each new spin being added with an identical localized wave function to the previous ones, mostly coupled to its own environment of nuclear spins. Projecting the sum of $2S$ such single-spin hyperfine terms ($\sum_{l=1}^{2S} \vec{S}_l\cdot\sum_{j=1} A_{l,j}\vec{I}_{l,j}$) to the fully symmetric subspace (total spin quantum number $S$) each hyperfine term contributes only with a fraction $1/(2S)$, giving rise to the factor in Eqs.~(\ref{fermi})-(\ref{dipolar}). In the other extreme, all the electrons couple in similar strength to all nuclei ($\sum_l \vec{S}_l\cdot \sum_j A_j I_j = \vec{S}\cdot \sum_j A^{(S)}_j I_j$), in which case projection to the symmetric subspace leads to no reduction with $S$. However, the $A_j$ themselves might be $S$-dependent. For instance, if they are contact terms that arise from an electron homogeneously delocalized over the $\sim (2S)^2$ nuclei, then we would see a quadratic decrease of the $A_j$. This is to say that beyond the $1/(2S)$ scaling both an additional reduction of the HFI with $S$ could arise (as the electron wave function spreads out over a larger number of sites) or it could increase  (as more nuclear spins contribute). As one might expect from the strong concentration of spin density and HFI on the edge sites of the molecule, the spin-chain picture describes the observed HFI well.

To see this, we will consider two specific numbers that can be unambiguously defined for the whole family of molecules. Namely the HFI of the maximally coupled nucleus of either species (for both species these are the one or two nuclei in the center of each edge of the triangle) and $A_{\mathrm{tot}}=\sum_j A_{zz,j}/2$ the sum of all $A_{zz,j}$ eigenvalues (divided by two). 
For hydrogen, this latter number represents the Overhauser shift that would be observed in a fully polarized \carbon-free molecule, while the sum of both terms is the Overhauser shift in an all-\carbon, fully polarized molecule. We plot the numbers for \hydrogen\ and \carbon\ separately since a different scaling behaviour might be expected (as the number of the former grows linearly and the latter quadratically with $n$). 

\begin{figure}
    \centering
    \includegraphics[width=\columnwidth]{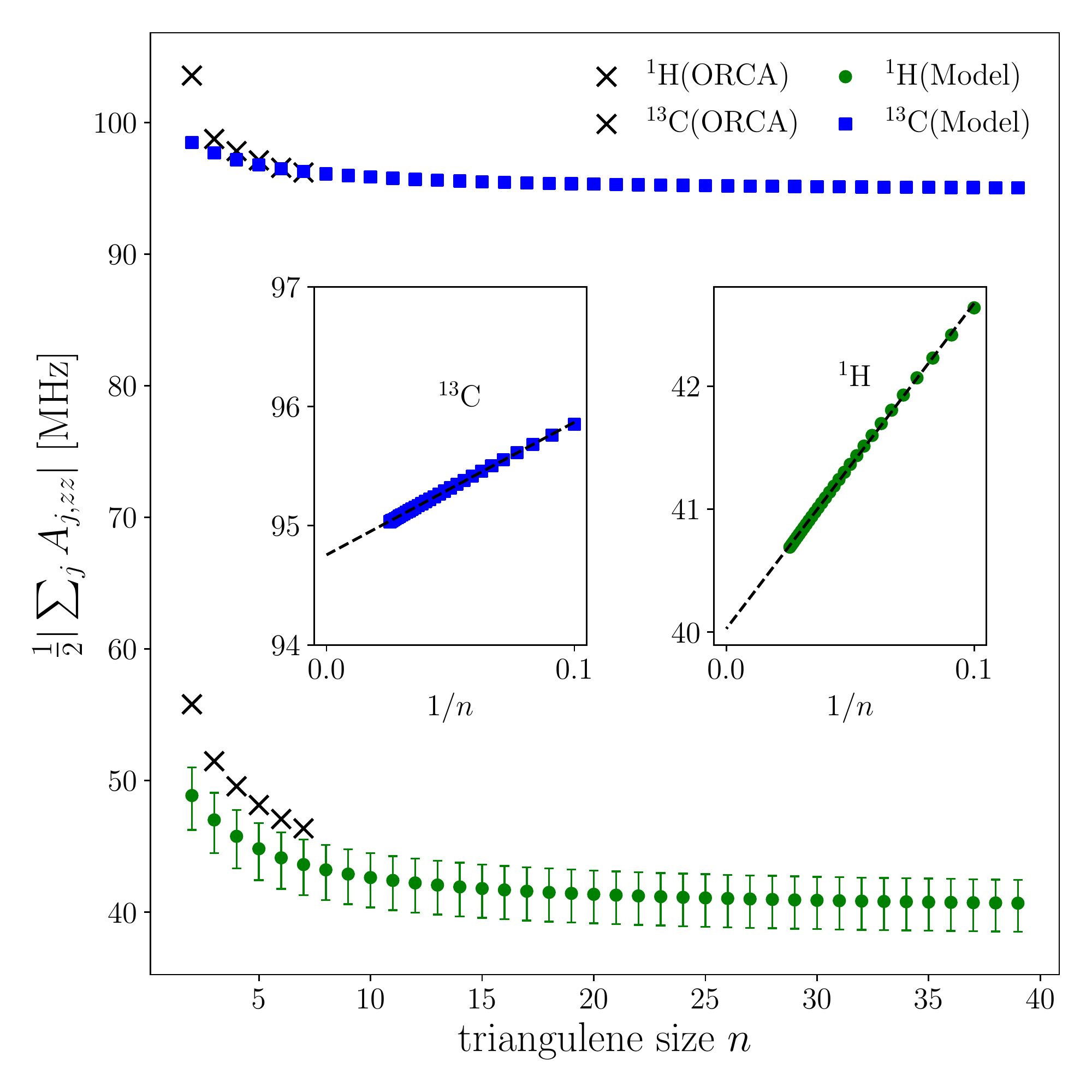}
    \caption{Characterization of HFI in a pure \carbon-[$n$]-triangulene.
    Inset shows the scaling behavior of large $n$ with saturation values at zero for \carbon\ and \hydrogen\ in the limit $n\rightarrow\infty$. 
    Crosses correspond to \orcaxc, while the filled symbols to MFH ($U=4$ eV). 
    Error bars reflect a $\pm$5$\%$ geometry scaling with respect to the non-scaled geometry leading to $\pm 4$ MHz variation in \hydrogen\ couplings. } 
    \label{fig:large-MFH}
\end{figure}

\begin{figure}
    \centering
    \includegraphics[width=\columnwidth]{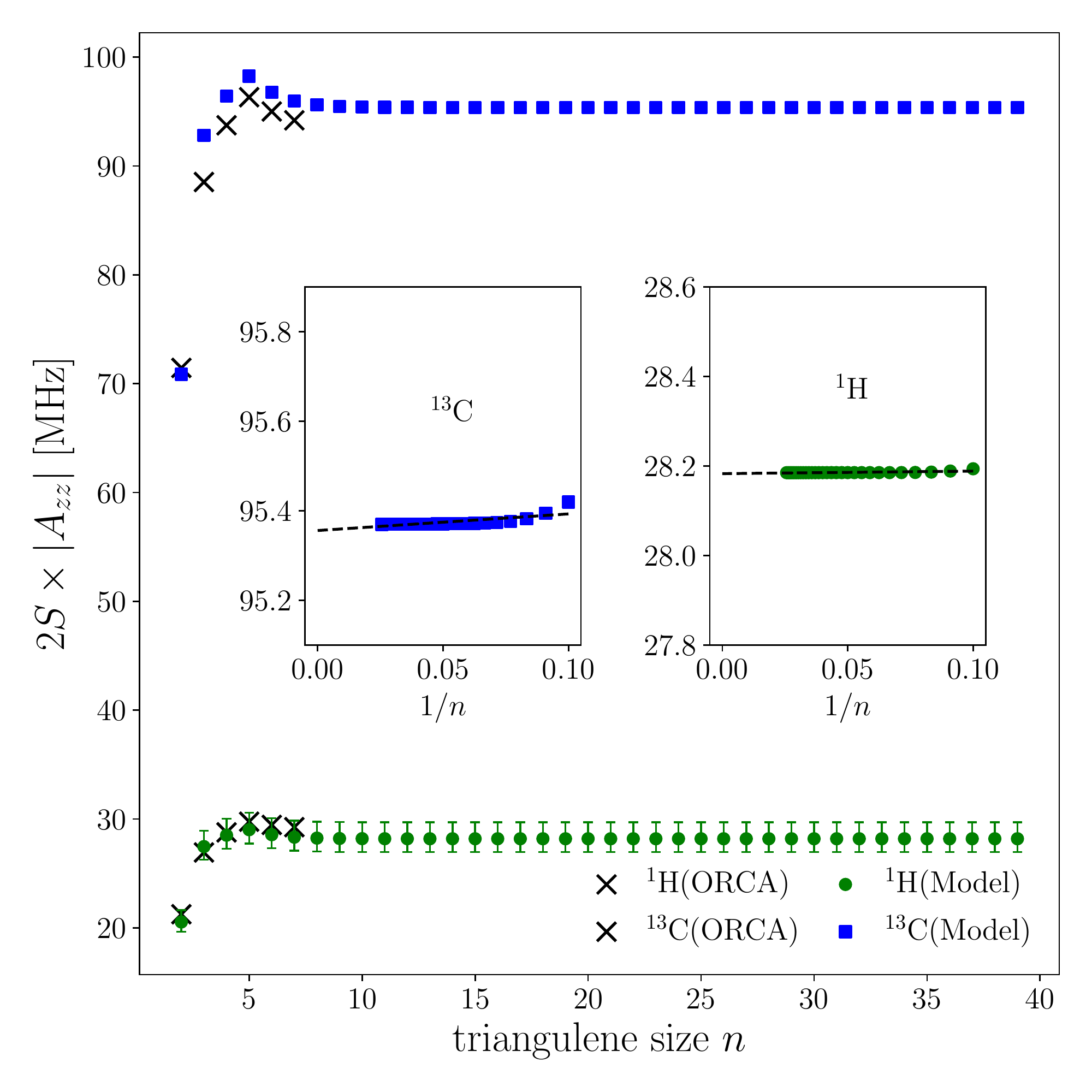}
    \caption{Largest $A_{zz}$ coupling (scaled by $2S$) for [$n$]triangulenes.
    Crosses correspond to \orcaxc, while the filled symbols to MFH ($U=4$ eV). 
    Error bars reflect geometry scaling of $\pm 5\%$ leads to variations in \hydrogen\ couplings within $\pm$ 4 MHz. } 
    \label{fig:large-abs-MFH}
\end{figure}

But, as we can see from \Figref{fig:large-MFH} there is no strong difference in the dependence of these totals on $n$: both decrease with $n$ in a way that for large $n$ becomes linear in $1/n$ and approaches a saturation value of $94$ MHz for \carbon\ and $41$ MHz for \hydrogen. At the same time, we see that the \emph{maximum} $A_{zz}$ hyperfine coupling decreases as $1/(2S)$ with $2SA_{zz}$ converging to 95 (29) MHz for \carbon\ (\hydrogen). 

That both numbers saturate show that all the electrons essentially localize in a region of finite width at the edges of the molecule with only a decreasing fraction of the overall spin density in the interior. For \hydrogen\, the numbers behave exactly as the \textit{spin-chain} model suggests that the maximum coupling decreases as $1/(2S)$. However, the sum converges to a constant that is given by $41 \mathrm{MHz}=|\sum_j A_{zz,j}|/2 \approx 3n/2\times(A_{zz}^{\mathrm{max}}/(2S)\approx43$ MHz, where the three appears because there are three additional strongly coupled hydrogens as $n$ is increased by 1. For carbon, a similar reasoning overestimates the sum, reflecting the importance of not just the maximally coupled nuclei at the edge of the molecule. 

Let us mention that we use an idealized geometry, not the relaxed one. Comparing to the DFT-relaxed geometries we see that (i) HFI changes little ($\sim\pm$ 4 MHz) and (ii) the distances change little (few $\pm 5\%$). To check the robustness of our numbers for the large molecules (which uses the idealized geometry) we then consider scaling by $\pm5\%$ and find again only small changes (Figs.~\ref{fig:large-MFH}, ~\ref{fig:large-abs-MFH}).

In the next section, we consider some paradigmatic applications of the hyperfine tensor by estimating the spin-qubit dephasing times $T_{2}^{*}$ for several of the considered molecules as well as sketching how HFI can be used to entangle a pair of distant nuclear spins.

\section{Applications of Hyperfine Tensor}
\label{sec:application}
\subsection{Dephasing of the electron spin}
In addition to its relevance for NMR experiments, knowledge of the hyperfine tensors allows further investigations. For example, in an external magnetic field oriented along $z$, the HFI component $A_{zz}$ to a random quasi-static effective magnetic field $h^{z}=\sum_j A_{zz,j} I^{z}_j$ experienced by the electron spin (Overhauser field). Precession in this unknown field leads to the dephasing of electron-spin coherence in a characteristic time, that we denote by $T_2^*$. Let us remark here that there are, in general, other processes contributing to the decay of spin-coherence, such as the transverse terms of the HFI, spin-orbit coupling, internal nuclear dynamics of magnetic-field noise that we do not consider. While the Overhauser-field fluctuations often lead to the fastest dephasing, this can, in principle, be fully reversed by spin-echo techniques \cite{abragam1961principles}. For identically and independently polarized static nuclear spins the coherence decays to good approximation according to a Gaussian decay law $\exp(-(t/T_2^*)^2)$ \cite{merkulov,coish}.

Assuming a polarization $P=\langle I^{z}_j\rangle$ for all nuclei, the variance of $h^{z}$ is given by 
\begin{equation}\label{variance}
    \sigma^2=\frac{1-P^2}{4}\sum_j A_{zz,j}^2,
\end{equation} 
which enables us to define
\begin{equation}\label{t2star}
    T_2^* = \frac {\sqrt{2}}{\sigma},
\end{equation}
denoting the time for which the electron-spin coherence decays to $1/e$.
While $T_2^*$ accurately describes the initial decay of electron spin coherence, we note that due to the small size of many of the nuclear spin baths considered here, deviations from Gaussian decay and, in particular, revivals of coherence can occur at larger times. 

Carbon-based materials show little HFI due to the low concentration of \carbon. But for the nanostructures we consider here, due to the large number of strongly coupled hydrogen nuclei even isotopically purified, \carbon-free molecules have a sizeable HFI and spin qubits will experience the corresponding dephasing. Thus, for example, in a molecule of [2]triangulene in a magnetic field in $z$-direction and with unpolarized nuclei, even if is \carbon-free, a dephasing time of $\sim53.5$ns can be expected due to the hydrogen nuclear spins, which is four times longer than the $T_2^*$ to be expected for an all-\carbon\ molecule. To reduce that interaction, it might be advantageous to identify electronic states for which the spin density is localized further away from hydrogenated edges. In \Tabref{tab:variance}, we give the magnitudes of $T_2^*$ for all the molecules investigated in this work. We mention these values both for pure $^{12}$C molecule with an hydrogen-only spin bath such that the dephasing time is given by $T_{2,\mathrm{H}}^{*}$ and for an all-\carbon\ molecule this is represented by $T_{2,\mathrm{all}}^{*}$.

\begin{table}
\begin{tabular}{lcrrrr}
\toprule
Molecule & $S$ 
    & $\sigma_{\mathrm{H}}$ & $T_{2, \mathrm{H}}^{*}$
    & $\sigma_{\mathrm{all}}$ & $T_{2, \mathrm{all}}^{*}$ \\
&& [MHz]  & [ns] & [MHz] & [ns] \\ 
\midrule
2T & 1/2              & 26.4 &  53.5      & 104.0 & 13.6 \\
3T & 1                & 18.7 &  75.6      &  75.6 & 18.7 \\
4T & 3/2              & 15.2 &  92.7      &  63.0 & 22.5 \\
5T & 2                & 13.1 & 108.2      &  54.9 & 25.8 \\
6T & 5/2              & 11.6 & 122.3      &  49.1 & 28.8 \\
7T & 3                & 10.5 & 134.9      &  44.9 & 31.5 \\
7AGNR$\dagger$ & 1/2  & 21.3 & 66.5       &  91.7 & 15.4 \\
indene & 1/2          & 26.2 & 53.9       &  92.2 & 15.3 \\
fluorene &  1/2       & 24.0 & 59.0       &  91.7 & 15.4 \\
\multicolumn{1}{p{20mm}}{indeno[2,1-b]-fluorene*} 
& 1                   & 16.7 & 84.8       &  66.1 & 21.4 \\
\bottomrule
\end{tabular}
\flushleft $\dagger$ topological end state;  *excited triplet state
\caption{\label{tab:variance} $\sigma_{\mathrm{H}/\mathrm{all}}$, square root of the variance, of the Overhauser field of unpolarized nuclei for pure-$^{12}$C and all-\carbon\ molecules, respectively, using \orcaxc. $T_{2,H}^{*}$ and $T_{2,\mathrm{all}}^{*}$ are the corresponding dephasing times for two adjacent Zeeman levels. Note that the molecules are always accompanied by H attached to C sites which gives rise to a H-spin bath.} 
\end{table}

\subsection{Electron-mediated long-range coupling of nuclear spins}
In sufficiently cold and clean systems, HFI can also be used to modify the nuclear state and dynamics via the electronic state. As an illustrative example we sketch how to obtain a direct interaction between distant nuclear spins. This electron-mediated interaction has been studied in detail for the case for electron-spin qubits in quantum dots, see, \eg, \cite{YLS06} can be much larger than the intrinsic nuclear interactions and can couple distant nuclear spins interacting with the same electronic system. As an exemplary case, we consider two carbon nuclear spins one at each edge of an AGNR and coupled to the electron spin in the end-state of the GNR (see \Figref{fig:hfi_spin_AGNR}). We consider a long GNR such that the singlet and triplet states are close in energy and apply an external magnetic field to bring one triplet state close to resonance with the singlet. The hybridization energy between the two end states decreases exponentially with the length of the AGNR. Using our tight-binding parametrization described above, we estimate that at a length of $15-20$ anthracene units ($\sim 6-8$nm) the singlet-triplet splitting is on the order of $100$ $\mu$eV, such that two of the states can be tuned into resonance with a magnetic field of $\sim2$ tesla. 
This is a situation very close to the one studied with a singlet-triplet qubit in a two-electron double quantum dots \cite{hanson}. We will therefore describe the electronic system as an effective two-level system, denoting by the corresponding spin operator by $\tilde{S}^z = (\proj{\text{singlet}}-\proj{\dn\dn})/2$ (see also \Appref{app:application}).

For simplicity of notation, we consider two symmetrically placed carbon nuclei so that they couple identically to the respective end state and we neglect hyperfine coupling to the other end state (which is very small due to the length of the GNR). To obtain the second-order interaction between the two nuclear spins we perform a Schrieffer-Wolff transformation \cite{BDL11}
with generator $T=\alpha \tilde{S}^+(I_1^+-I_2^+) + \beta \tilde{S}^+(I_1^--I_2^-) - \mathrm{h.c.}$, where 
$\beta=-(A_{xx}+A_{yy})/(2\sqrt{2}\tilde{\omega})$ and $\alpha=(A_{xx}-A_{yy})/(2\sqrt{2}\tilde{\omega})$ and $\tilde{\omega}=\omega-\Delta$ is an effective Zeeman splitting given by the difference between the electronic Zeeman splitting $\omega$ and the singlet-triplet splitting $\Delta$, cf.~\Appref{app:application}. We obtain an effective nuclear Hamiltonian in the low-energy subspace given by 
\begin{equation}\label{eq:nncoupling}
H_{\mathrm{eff}} = \nu (I_1^z+I_2^z)+ J_x I_1^xI_2^x+ J_y I_1^y I_2^y,
\end{equation}
describing two spins with (renormalized) Zeeman splitting and a mediated $XY$ interaction term between the two nuclei at the far ends of the GNR. The coupling constants $J_{x,y}$ are found to be $J_{u}=
A_{u}^{2}/\tilde{\omega}-(A_{x}-A_{y})^{2}/(2\tilde{\omega})$ for $u=x,y$. According to our HFI calculations the $J_u$ could be on the order of several MHz, which is orders of magnitudes larger than the intrinsic nuclear dipole-dipole coupling even between neighboring carbon spins. We give some values for the coupling constant $J_u$ for the 7AGNR. In presence of a perpendicular magnetic field, typical values of in-plane HFIs are on the order of $|A_{x}| \sim 11.3$ MHz, $|A_{y}|\sim 13.2$ MHz and $\tilde{\omega} \sim 50$ MHz which leads to $J_x \sim 2.5$ MHz. Since for our system the out-of-plane eigenvalue $A_{zz}$ can be much larger than the in-plane eigenvalues of $A$, the coupling can also be enhanced by applying the magnetic field in plane, say, in $x$-direction. Then in the above equations the $A_{u}$ have to be permuted cyclically, \ie, the coupling is obtained from  $|A_z| \sim 38.1$ MHz and $|A_y| \sim 19.2$ MHz, leading to $J_z \sim 25.5$ MHz.

For external fields of $\sim1-2$ tesla as needed to bring singlet and triplet close to resonance, we find that the nuclear Zeeman energies $\nu$ are of order 10-20 MHz for \carbon\ (and four times that for hydrogen), \ie, significantly larger than the mediated coupling. Consequently the interaction will be dominated by the $I^z$-conserving exchange term $\propto I_1^+I_2^- + I_1^-I_2^+$ with strength $(J_x+J_y)/2$.

\section{Discussion and Outlook}
\label{sec:discussion}

In summary, we have investigated the hyperfine interactions of open-shell planar $sp^2$-carbon nanostructures belonging to the class of benzenoid as well as non-benzenoid molecules. Within first-principles calculations using DFT as implemented in \orca, we have found that both isotropic Fermi contact and anisotropic terms contribute significantly to the overall hyperfine couplings. The sizable Fermi contact terms demonstrate/confirm the importance of polarization of core electrons and $\sigma$ states \cite{Ya.08.HyperfineInteractionsGraphene, KaFr.61.TheoreticalInterpretationCarbon13}. We find that in addition to \carbon\ the \hydrogen\  nuclei play an important role for HFIs in these molecules.
This implies that even in pure-$^{12}$C molecules, HFI is non-negligible (and not much weaker than for natural abundance of \carbon). As some of these molecules exhibit high-spin ground states, we have also explored their effects on HFIs. Generally, the magnitude of HFIs is seen to decrease with the increase in the number of unpaired electrons and thus for higher spin states, HFIs are typically smaller.
The largest HFI for \carbon\ in the benzenoid structures studied is found to be around 90 MHz for a spin doublet (namely the 7AGNR), whereas for the non-benzenoid case the maximum HFI appears on the pentagons with a magnitude $\sim$ 120 MHz. Likewise for \carbon-free molecules, the largest magnitude of HFI for \hydrogen\ in benzenoid and non-benzenoid molecules is found to be $-37$ MHz (for 7AGNR) and $-62$ MHz (for fluorene), respectively. 

We find that the calculated HFIs in the molecules we studied track closely the atom-resolved Mulliken \pol\ at the carbon sites. Exploiting this relation, we provide generic $sp^2$-carbon fit parameters for HFIs for both the \carbon\ and \hydrogen\ nuclei. They are obtained by extending the fitting procedure of Refs.~\cite{KaFr.61.TheoreticalInterpretationCarbon13, McCh.58.TheoryIsotropicHyperfine, Ya.08.HyperfineInteractionsGraphene}. Our results imply that both on-site and nearest-neighbor carbon \pol\ play a role in the Fermi contact as well as dipolar HFIs. Once the carbon \pol\ are known using a suitable electronic structure method, one can use these generic $sp^2$ fit parameters to obtain the HFIs for any molecule belonging to this group, without having to perform real-space integrals and without having to involve $s$ electrons. 

Moreover, our work shows that simple approaches based on \siesta\ or MFH when combined with the empirical parametrization procedure can provide a valuable alternative toolkit to compute hyperfine tensors for these classes of molecules. Using MFH, we report on the scaling of the HFI with triangulene size which cannot be afforded within the \orca\ implementation, allowing us to compute thousands of atoms within minutes on a standard laptop without the use of supercomputers.
Our results indicate that with increasing system sizes, the Overhauser field for an all-\carbon, fully polarized molecule saturates to $\sim$ 95 MHz, while for a purely-$^{12}$C molecule, that value would be $\sim 41$ MHz.
For this, we performed MFH calculations with up to 1678 carbon sites.
Additionally, the fitting procedure opens a possibility to study HFI in physical systems that require periodic boundary conditions (Bloch's theorem) \cite{soler} or infinite, non-periodic boundary conditions (Green's functions) \cite{PaLoFr.17.Improvementsnonequilibrium, SaPaGi.22.SpinPolarizingElectron}. 
Furthermore, it would be interesting to explore whether the fitting procedure can be extended to magnetic hydrocarbons incorporating certain heteroatoms (like B or N, see \cite{OtFr.22.Carbonbasednanostructures} and non-planar molecules (\eg, oligo(indenoindene) chains \cite{ortiz}).

In this work we have only considered HFI in \emph{isolated} molecules and its corresponding relations with \pol\ at the carbon sites. 
However, often $\pi$-magnetic nanostructures are fabricated and characterized on noble metal surfaces -- like Au(111) -- where the molecular states hybridize significantly with the substrate states \cite{OtFr.22.Carbonbasednanostructures}.
This raises a question on how electronic hybridization affects the HFI within the nanostructure.
Furthermore, on insulating substrates and thin films -- like MgO \cite{BaPaCh.15.Electronparamagneticresonance, wilke}, NaCl \cite{repp22}, 
TiO$_2$ \cite{KoStIz.20.Rationalsynthesisatomically} -- one may ask about modifications to the intrinsic HFI due to chemical bonding \cite{ShDiGu.21.Trendshyperfineinteractions} as well as to the adjacent nuclear spins in the substrate.

The larger values of HFIs for \carbon\ as well as \hydrogen\ in these molecules have significant implications. Firstly, in terms of experimental verification of HFIs, these molecules could serve as candidates in STM-ESR techniques \cite{wilke}. The ESR spectrum could be notably different for a pure $^{12}$C molecule than for one with 1$\%$ \carbon\ nuclei. While in the former, the spectrum would be dominated by the large HFIs of \hydrogen\ nuclei, the latter would show appreciable signatures of the \carbon\ nuclei if they are located at one of the strongly coupled sites.
Secondly, they have an observable influence on the electronic spin coherence time. In a magnetic field of the order of few millitesla, there is no spin exchange between electron and nuclei due to the vastly different energy scales and the HFIs lead to a random and fluctuating effective magnetic field (the Overhauser field) experienced by the electron spin. The effect of this field on the electron spin coherence can then be gauged by the $T_{2}^{*}$ dephasing times \cite{merkulov,coish}. For all the molecules investigated in this work, we have provided the dephasing times both for an all-\carbon\ molecule as well as pure $^{12}$C molecule (\Tabref{tab:variance}). Additionally, HFI  gives rise to an (electron-mediated) nuclear-spin dynamics that causes electron spin decoherence. We leave it as future work to explore the quantitative aspects of spin decoherence in these molecules which requires a treatment of internal nuclear dynamics and electron-nuclear entanglement.

It is well known that effects of HFIs can go beyond the decoherence of localized electron spins. All the molecules studied here are different realizations of a central-spin system in which one spin-$S$ is coupled to many spin-$I$s (by the hyperfine tensor $A_j$ for the $j$th spin $I_j$). For mutually commuting $A_j$ this gives rise to the integrable Gaudin magnet \cite{Gaudin1976, Dukelsky2004}, a paradigmatic spin model that has been used to study decoherence \cite{Prokofev2000}, quantum \cite{Shao2023} and dissipative \cite{Kessler2012a} phase transitions, as well as time crystals \cite{Randall2021,Frantzeskakis23}. For the systems studied here, the $A_j$ do, generally, not commute. While the integrability is lost for general $A_j$, the characteristic one-to-all connectivity of the central spin remains. The symmetric nature of many of the molecules studied can give rise to a \emph{structured} spin bath that may lead to pronounced revivals, another signature of quantum-coherence in these nanosystems \cite{Heinrich2021}. 

With a view to the possible realization of spin qubits in these structures, we first remark that the requisite coherent control over single spins has not been demonstrated so far in the systems we study, although the STM- and AFM-ESR techniques demonstrated recently \cite{BaPaCh.15.Electronparamagneticresonance,wilke,repp22} put the on-surface control of atomic-scale spins within reach.
As mentioned, our results show that electron spins in graphene nanostructures have to contend with a nuclear-spin bath leading to dephasing times on the scale of 10-100 ns. While this is much longer than the timescale of electron-spin interactions in these systems (ps for meV interactions \cite{OtFr.22.Carbonbasednanostructures}), the strength and always-on nature of these interactions may make the direct use of electron-spin qubits challenging.
However, the appearance of few, strongly and inhomogeneously coupled nuclear spins makes these systems very suitable to employ the nuclear spins as the primary quantum memory and register (this approach has been pursued for several spin-qubit platforms \cite{Wolfowicz2020} and demonstrated most compellingly for nuclear spins surrounding nitrogen-vacancy (NV) centers \cite{Abobeih2019,Bradley2019}), while using the interacting electron system to mediate interaction between distant nuclear spin qubits.
As we sketched in the previous subsection, our results for the HFI coupling strengths indicate that interactions on the MHz scale between distant nuclei may be achieved. It will be interesting to study this for larger systems of interacting electron spins, \eg, triangulene chains \cite{Mishra2021} or GNRs hosting spin chains at their edge \cite{Sun2020}, would allow to exploit strong, long-range electron-spin interactions to realize quantum gates between isolated nuclear spins. The graphene nanostructures studied here combine advantages of NV centers (resolvable nuclear-spin bath) and quantum-dot arrays (strong electronic interactions). 

It has been shown that ESR and NMR techniques combined with the interactions of the central-spin Hamiltonian allow for universal control of the electron-nuclear spin system \cite{Casanova2017}. To assess whether these results could lead to practical protocols for the systems considered here, further studies, in particular regarding the robustness to various sources of noise (presence of a nuclear spin bath, electron tunneling, spin-orbit interaction) and the achievable control over electron and nuclear spin states and dynamics need to be explored.

\section*{Acknowledgments}
This work was supported by the European Union (EU) through Horizon 2020 (FET-Open project SPRING Grant no.~863098) and the Spanish MCIN/AEI/ 10.13039/501100011033 (PID2020-115406GB-I00).

\appendix

\section{Orientation of hyperfine eigenvectors}
\label{app:vec_2t}

We provide here a brief qualitative and quantitative understanding of the eigenvectors shown in \Figref{fig:2Teig} for 2T, where we note that for certain atomic positions, the axes are noticeably rotated compared to the bonds. For the carbon atoms, where a bond corresponds to a symmetry axis of the molecule, there is an eigenvector along that bond (for the central nucleus C4 the in-plane eigenvalues are degenerate and hence the direction arbitrary). For the majority-site carbons (C2 and equivalent) no such symmetry exists and the eigenvectors are tilted.

\begin{figure*}[h]
    \centering
    \includegraphics[width=0.9\textwidth]{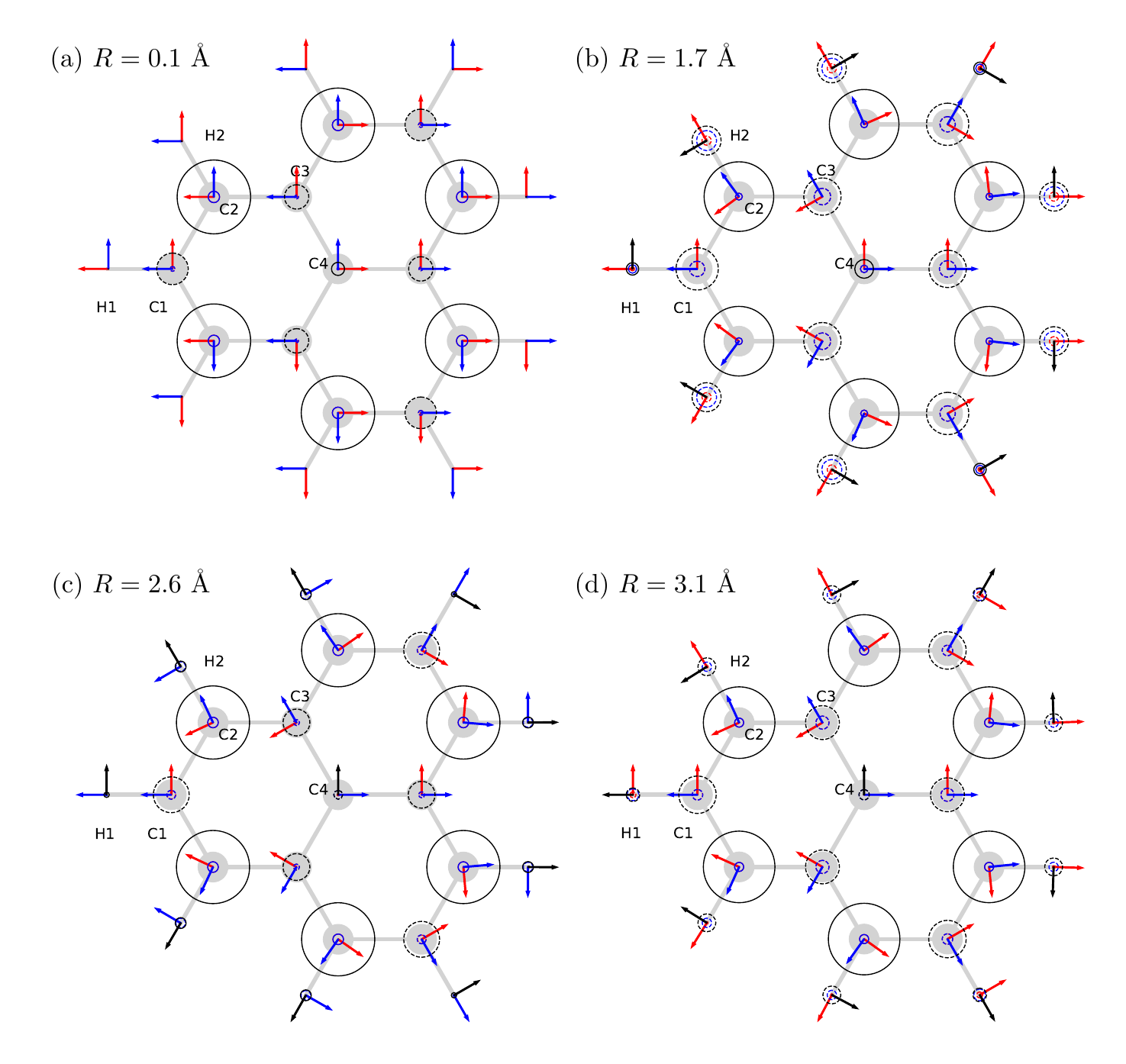}
    \caption{\label{fig:hfi_vectors}
    Orientation of hyperfine eigenvectors for 2T considering different interaction radius $R$ for the dipolar hyperfine fields in the fitting procedure for the HFI tensors. The panels correspond to the inclusion of (a) no neighbor sites ($R=0.1$ \AA), (b) nearest neighbor sites only ($R=1.7$ \AA), (c) up to second-nearest sites ($R=2.6$ \AA), and (d) up to third-nearest sites ($R=3.1$ \AA).
    In each case optimized fitting parameters against the 2T HFI tensors from \orcaxc\ were used.}
\end{figure*}

In \Figref{fig:hfi_vectors} we show how this emerges in our fitting procedure for 2T as the number of neighbor sites (shells) is varied: Using the ORCA-derived HFI tensor for 2T as reference in each case, we obtain fitting parameters (similar to those given in \Secref{sec:3-fitting-hfi}) and the corresponding fitted HFI tensors by varying the radius of interaction $R$, centered around an on-site atomic position. 
If one takes into account only the on-site \pol, as shown in \Figref{fig:hfi_vectors}(a) for $R=0.1$ \AA, the in-plane eigenvalues are degenerate and the orientation of the corresponding vectors is arbitrary (in this case along the $xy$ coordinate axes).
When first-nearest neighbor sites (as considered in the main text) are included in the fitting, as shown in \Figref{fig:hfi_vectors}(b) for $R=1.7$ \AA, one in-plane vector generally orients along a bond. The exception to this are the vectors for the C2 atoms, which display a slight rotation away from the bond axis. This asymmetry can be understood from the fact that the {\pol}s at sites C1 and C3 are not identical.
When we increase further to $R=2.6$ \AA, as shown in \Figref{fig:hfi_vectors}(c), contributions from up to second-nearest neighbors are included. This leads to significant changes in the eigenvalues and the slight rotation angle of the C2 vectors are now in the opposite direction with respect to the bond.
Finally, including up to third-nearest neighbors with $R=3.1$ \AA, as shown in \Figref{fig:hfi_vectors}(d), the orientation of the eigenvectors is essentially unchanged, but the eigenvalues are slightly modified.
Compared with the reference ORCA calculation shown in \Figref{fig:2Teig}, 
we thus conclude that the deviation of in-plane eigenvectors from the bond axis is indicative of the asymmetric contribution of the second- and third-nearest neighbor dipolar hyperfine fields.

We note that within our model, defined by Eqs.~(\ref{model:fermi})- (\ref{model:dipolar}), there is no systematic improvement by increasing $R$. In fact, the best fit (in terms of RMSE) is obtained for $R=1.7$ \AA. 
If one were to introduce independent fitting parameters for the different shells (as opposed to a common one as in our treatment) an even better fit could probably be obtained.

\section{Electron-nuclear Hamiltonian}
\label{app:application}

We start from the hyperfine Hamiltonian of two nuclear spin-1/2 $I_{j}, j=1,2$ each coupled with hyperfine tensor $A_j$ to an electron spin-1/2, while the two electrons
are exchange-coupled. For clarity of the resulting expressions, we assume the the two hyperfine tensors are identical, which would be the case, \eg, for nuclei at
symmetric positions if one of their in-plane eigenvectors points along the ribbon, and we apply an external magnetic field along one of the three
eigendirection of the hyperfine tensor, which we call $z$ in this appendix even though it need not be perpendicular to the plane of the molecule.

The relevant spin Hamiltonian is then given by
\begin{equation}
H = \omega (S_{1}^z+S_{2}^{z}) -\Delta\proj{S}+ \nu (I_{1}^{2}+I_{2}^{z}) +
  \bS_{1}A_{1}\bI_{1}+ \bS_{2}A_{2}\bI_{2},
\end{equation}
where $\ket{S}$ denotes the electronic singlet state and $\Delta$ is the singlet triplet splitting. (Since $\bS_{1}\cdot\bS_{2}=\id/4-\proj{S}$, we could
also write $+\Delta \bS_{1}\cdot\bS_{2}$ up to a shift in energy.) It is convenient to also introduce the sums and differences of the spin operators $\bS=\bS_{1}+\bS_{2}, \bR=\bS_{1}-\bS_{2}$
for the electron spins and $\bI=\bI_{1}+\bI_{2}, \bK=\bI_{1}-\bI_{2}$ for the
nuclei. The Hamiltonian then contains a diagonal part $H_{d}=\omega S^{z}+\nu
I^{z}+\frac{A_{z}}{2}(S^{z}I^{z}+R^{z}K^{z})$ and the off-diagonal part
$H_{o}=\frac{A_{x}+A_{y}}{4}(S^{+}I^{-}+R^{+}K^{-}+\mathrm{h.c.})+\frac{A_{x}-A_{y}}{4}(S^{+}I^{+}+R^{+}K^{+}+\mathrm{h.c.})$. All
terms in $H_{o}$ involve transitions between electronic states that differ by an energy large compared to the hyperfine coupling, \ie, they are
off-resonant. They can be removed by a Schrieffer-Wolff transformation \cite{BDL11} (or quasidegenerate perturbation theory \cite{ShRe80}), which yields an effective Hamiltonian (block-diagonal in the
$S^{z}$ basis) that reveals the second-order (in $A_{j}/\omega$) electron-mediated interaction
between the nuclear spins. We focus here on the special case that the energy difference between the singlet and the low-energy triplet is much smaller than the Zeeman splitting but still much larger than the hyperfine interaction: $\omega_{Z}\gg |\omega-\Delta|/2\gg A_{j}$ (which can readily be realized as long as $\Delta$ is at least two orders of magnitude larger than the $A_{j}$). Then the second-order contributions will be dominated by those arising from coupling between just
the two states (singlet and low-energy triplet) close in energy and we can treat the electron spins as an effective two-level system spanned by the states
$\ket{0}=\ket{S}$ and $\ket{1}=\ket{\dn\dn}$ with effective Zeeman splitting $\tilde{\omega}=(\omega-\Delta)$
and perform the Schrieffer-Wolff transformation for the simplified Hamiltonian
\begin{equation}\label{eq:app:hfi}
\begin{split}
  \tilde{H} = &\, \tilde{\omega}\tilde{S}^{z}+\tilde{\nu} I^{z} +
  \frac{A_{z}}{2}\tilde{S}^{z}I^{z}+\frac{\bar{A}}{\sqrt{2}}(\tilde{S}^{+}K^{-}+\mathrm{h.c.})\\
  &+\frac{\Delta
    A}{\sqrt{2}}(\tilde{S}^{+}K^{+}+\mathrm{h.c.}),\\
\end{split}
\end{equation}
where we have introduced $\tilde{S}^{z}=(\proj{1}-\proj{0})/2$, $\tilde{S}^{+}=\ketbra{1}{0}$ and $\tilde{\nu}=\nu-A_{z}/4$ and $\bar{A}=(A_{x}+A_{y})/2, \Delta A=(A_{x}-A_{y})/2$. Changing basis with
$U=\exp(T)$, where 
\begin{equation}
  \label{eq:1}
  T=\alpha \tilde{S}^{+}K^{+}+\beta \tilde{S}^{+} K^{-}-\mathrm{h.c.}
\end{equation}
and $\alpha=\Delta A/(\sqrt{2}\tilde{\omega})$ and $\beta=\bar{A}/(\sqrt{2}\tilde{\omega})$ are
chosen such that $[T,\tilde{\omega}\tilde{S}^{z}]=-H_{o}$, 
we obtain
    \footnote{We use $[T,\tilde{S}^{+}K^{-}+\tilde{S}^{-}K^{+}] = -4\beta
    \tilde{S}^{z}-2\beta
    I^{z}+4\beta\tilde{S}^{z}(I_{1}^{+}I_{2}^{-}+h.c.)-4\alpha\tilde{S}^{z}(I_{1}^{+}I_{2}^{+}+h.c.)$ and $[T,\tilde{S}^{+}K^{+}+\tilde{S}^{-}K^{-}]=4\alpha \tilde{S}^{z}+2\alpha
    I^{z}-4\alpha\tilde{S}^{z}(I_{1}^{+}I_{2}^{-}+\mathrm{h.c.})-4\beta\tilde{S}^{z}(I_{1}^{+}I_{2}^{+}+\mathrm{h.c.})$. The commutators with the other diagonal terms lead to higher-order off-diagonal terms which we neglect here and which can be removed by adding higher-order terms to $T$.}
--in the low-energy subspace and up to second order in 
$A_{j}/\tilde{\omega}$-- the block-diagonal Hamiltonian $H'=UHU^{\dag}$
\begin{equation}
  \label{eq:5}
\begin{split}
  H' =&\, \tilde{\omega}\tilde{S}^{z}+\nu'
       I^{z}+\frac{\bar{A}^{2}-(\Delta A)^{2}}{\tilde{\omega}/2}\tilde{S}^{z}\left(I_{1}^{+}I_{2}^{-}+I_{1}^{-}I_{2}^{+}\right)\\
  &-
  \frac{2\bar{A}\Delta A}{\tilde{\omega}/2}\tilde{S}^{z}\left(I_{1}^{+}I_{2}^{+}+I_{1}^{-}I_{2}^{-}\right)
\end{split}
\end{equation}
with the renormalized frequency $\nu'=\tilde{\nu}-(\bar{A}^{2}-(\Delta
A)^{2})/\tilde{\omega}$. Thus, for the electron in the singlet state $\ket{1}$ we obtain the effective
nuclear Hamiltonian 
\begin{align}
  \label{eq:7}
  \tilde{\nu}'I^{z}+ J_{x}I_{1}^{x}I_{2}^{x}+J_{y}I_{1}^{y}I_{2}^{y}
\end{align}
with $J_{x/y}=\frac{\bar{A}^{2}+(\Delta A)^{2}+/-2\bar{A}\Delta
  A}{\tilde{\omega}/2}$, \ie, $J_{u}=
A_{u}^{2}/\tilde{\omega}-(A_{x}-A_{y})^{2}/(2\tilde{\omega})$ for $u=x,y$.

\clearpage
%

\end{document}